\documentclass{aastex631}
\usepackage{empheq}

\usepackage{xcolor}
\usepackage{import}
\usepackage[caption=false]{subfig}

\newcommand{\HII}{\ion{H}{2} }	
\newcommand{\OIII}{[\ion{O}{3}] }	
\newcommand{\OII}{[\ion{O}{2}] }	
\newcommand{\HA}{H$\alpha$ }			
\newcommand{\fluxunitIV}{$\times\ 10^{-14}$ erg s$^{-1}$ cm$^{-2}$}
\newcommand{\fluxunitV}{$\times\ 10^{-15}$ erg s$^{-1}$ cm$^{-2}$}
\newcommand{\fluxunitVI}{$\times\ 10^{-16}$ erg s$^{-1}$ cm$^{-2}$}
\newcommand{\logfluxunit}{ erg s$^{-1}$ cm$^{-2}$}

\shorttitle{SFACT II}
\shortauthors{Sieben et al.}
\graphicspath{{./}{./}}

\begin{document}

\title{The Star Formation Across Cosmic Time (SFACT) Survey. II. The First Catalog from a New Narrow-Band Survey for Emission-Line Objects}



\author[0000-0002-5513-4773]{Jennifer Sieben \footnote{Visiting astronomer, Kitt Peak National Observatory at NSF’s NOIRLab, which is managed by the Association of Universities for Research in Astronomy (AURA) under a cooperative agreement with the National Science Foundation.}}
\affiliation{Indiana University, 727 East 3rd Street, Bloomington, IN 47405 USA}

\author[0000-0002-4876-5382]{David J. Carr}
\affiliation{Indiana University, 727 East 3rd Street, Bloomington, IN 47405 USA}

\author[0000-0001-8483-603X]{John J. Salzer}
\affiliation{Indiana University, 727 East 3rd Street, Bloomington, IN 47405 USA}

\author[0000-0002-2954-8622]{Alec S.\ Hirschauer}
\affiliation{Indiana University, 727 East 3rd Street, Bloomington, IN 47405 USA}
\affil{Space Telescope Science Institute, 3700 San Martin Drive, Baltimore, MD 21218, USA}

\begin{abstract}
Star Formation Across Cosmic Time (SFACT) is a new narrow-band survey designed to detect faint emission-line galaxies and QSOs over a broad range of redshifts. Here we present the first list of SFACT candidates from our pilot-study fields. Using the WIYN 3.5m telescope, we are able to achieve good image quality with excellent depth and routinely detect ELGs to r = 25.0.  The limiting line-flux of the survey is $\sim$1.0 \fluxunitVI .
SFACT targets three primary emission lines: H$\alpha$, \OIII $\lambda$5007, and \OII $\lambda$3727. The corresponding redshift windows allow for the detection of objects at $z\sim 0-1$. With a coverage of 1.50 deg$^2$ in our three pilot-study fields, a total of 533 SFACT candidates have been detected (355 candidates deg$^{-2}$). We detail the process by which these candidates are selected in an efficient and primarily automated manner, then tabulate accurate coordinates, broad-band photometry, and narrow-band fluxes for each source.

\end{abstract}


\section{Introduction} \label{sec:intro}

The Star Formation Across Cosmic Time (SFACT) survey is an ongoing wide-field imaging and spectroscopic program which targets the detection of large numbers of extragalactic emission-line sources. As a narrow-band (NB) survey, SFACT is able to discover a wealth of new sources exhibiting strong emission lines. The SFACT survey methodology draws upon the rich legacy of previous emission-line galaxy (ELG) surveys 
\citep[e.g.,][]{MacAlpine77,Markarian,BST1993, KISS1,rw2004, Kakazu2007, werk2010, ly2011, hadot1, HIZELS12, HIZELS13, Stroe2015, cook2019, hadot2, LAGER20, hadot4, miniJPAS}. SFACT builds on this previous work, using a medium-class telescope with a wide field of view and three custom NB filters. 

The goal of the SFACT survey is to produce a high quality catalog of emission-line objects whose selection function and completeness limits can be accurately quantified, so that the resulting catalog of ELGs will be useful for a broad range of studies requiring statistically-complete galaxy samples.   A comprehensive description of the survey is given in \citet[][henceforth referred to as SFACT1]{SFACT1}.   SFACT is designed to both cover a wide area on the sky and to be deep.   When completed, the total area covered by the survey  will be between 25 and 30 deg$^{2}$.  Furthermore, our data routinely reach a limiting line flux detection level of $\sim$1.0 \fluxunitVI .   These survey parameters represent compromises.   SFACT does not reach the ultra-faint flux levels of extremely deep NB surveys \citep[e.g.,][]{HIZELS12, HIZELS13, LAGER20}, but it does cover much larger survey areas.   Conversely, SFACT does not have the extreme FOV coverage of surveys like \citet{cook2019} and \citet{miniJPAS}, but it reaches to substantially deeper detection limits.

The current paper is one of a series of three initial SFACT publications that present survey results for our pilot-study fields.  SFACT1 presents the survey description, goals and motivation.   It also provides a summary of the properties of the 533 ELGs detected in our first three survey fields (magnitudes, luminosities, redshifts, line fluxes, star-formation rates, etc.).  Example objects are shown which illustrate the types of objects being detected by SFACT;  both imaging and spectroscopic data are presented.   SFACT1 also explores numerous sciences applications that can be addressed with the full survey.  The current paper (SFACT2) presents the initial survey lists selected from the imaging portion of the survey. We provide details of our observing and image processing procedures as well as how the ELGs are selected.   We illustrate our survey method with numerous examples of sources discovered in our NB images, and summarize the properties of the sample derived from our imaging data.  The third SFACT paper (\citealp{SFACT3}, henceforth referred to as SFACT3) focuses on the spectroscopic component of the survey, discussing the procedures for the observations and processing of the spectral data.   SFACT3 tabulates key spectroscopic data obtained for the ELGs in our pilot-study fields.   These data are used to verify the nature of the objects discovered in the imaging data and to derive a range of key parameters.  SFACT3 also presents the spectra corresponding to the example images shown in SFACT2.

In this paper, we first describe our observational procedures (Section~\ref{ssec:obs}) and our data processing technique (Section~\ref{ssec:proc}). Our method for selecting objects for inclusion in our survey catalogs is detailed in Section~\ref{Sec:Target selection}, along with our photometry method and calibration in Section~\ref{sec:photometry}. The results of the pilot study, including the data and example objects, are presented in Section~\ref{Sec:results}. For all of the SFACT papers we assume a standard $\Lambda$CDM cosmology with $\Omega_m$  = 0.27, $\Omega_\Lambda$ = 0.73, and H$_0$ = 70\ km\ s$^{-1}$\ Mpc$^{-1}$.

\section{Observations \& Data Processing}\label{Sec:ObsDP}
\subsection{Observations}\label{ssec:obs}
All survey imaging data were acquired using the One Degree Imager (ODI; ~\citealp{ODI}) on the WIYN\footnote{The WIYN Observatory is a joint facility of the University of Wisconsin-Madison, Indiana University, NSF's NOIRLab, the Pennsylvania State University, and Purdue University.} 3.5m telescope sited at Kitt Peak, Arizona. ODI consists of 30 Orthogonal Transfer Array (OTA) CCDs, each of which comprises 64 480 $\times$ 494 pixel cells. The pixel size for the ODI OTAs is 12$\mu$, which yields an image scale of 0.11\arcsec\ pixel$^{-1}$. The total field of view of ODI is 40\arcmin $\times$ 48\arcmin.   All survey fields are observed through six filters: three BB filters (gri) and three NB filters. The BB data were obtained through g, r, and i filters $\sim$1500~\AA \ in width. The BB bandpasses mimic the SDSS filters \citep{SDSS}. 

The fields observed for SFACT were selected to overlap with the Sloan Digital Sky Survey (SDSS,~\citealp{SDSS, SDSSDR15}), which we used for photometric calibrations. Two of the fields presented in this pilot study were centered on ELGs found in the previous \HA Dots survey \citep{hadot1, hadot2, hadot4}.  This provided a valuable testbed for the current survey methodology. The selection of the SFACT survey fields is discussed in more detail in SFACT1.

\subsubsection{Science Observations}
The imaging data used for this paper were obtained during three observing seasons. For a full list of observing dates, see Table~\ref{tab:obs dates}. In November of 2016, initial test data were acquired for the SFF10 and SFF15 fields in the r-band and first narrow-band filter (NB1). These observations provided the  data used to develop our processing and object-selection methods (see Section~\ref{Sec:Target selection}). In 2017 we added additional broad-band (BB) observations of SFF10 and SFF15 in g- and i-band plus included an additional field (SFF01). Our data set for the pilot study was then completed upon the subsequent addition of two additional NB filters in 2018. 

The NB data were obtained through three special filters designed for the survey, centered at 6590~\AA, 6950~\AA, and 7460~\AA, each with a width of $\sim$ 90~\AA\ (henceforth NB2, NB1, and NB3, respectively). The exact bandpasses are detailed in SFACT1 as well as the redshift ranges accessible via commonly detected emission lines. The transmission curves of our NB filters are shown in Figure~\ref{fig:filters}. The three NB filters fall within the r or i BB filters and are in a region where the CCD sensitivity is quite high. 

All NB and BB images were taken using a nine-point dither pattern. The dither sequence is a carefully-planned sequence of position adjustments in order to move sources off of bad columns, chip gaps, or dead OTA cells on the camera. By moving the telescope such that inactive areas on the camera are not always covering the same region on the sky, we ensured that we were truly covering the full available field of view. In this way, multiple exposures of the same fields increased image depth, allowing for the detection of fainter sources.

Each individual NB exposure was 600 seconds, for a total integration time of 90 minutes for each NB dither sequence. Because each pixel in the final stacked images is typically illuminated by the sky in only 6-7 images in the dither sequence, the effective exposure time is closer to 60-70 minutes for each pixel in the NB images. Each individual BB exposure was 120 seconds, and the final stacked BB images likewise include light from 6-7 images in a given pixel of the final stacked image.

\begin{deluxetable}{ccccccc}\label{tab:obs dates}
\tabletypesize{\footnotesize}
\tablecaption{Observation Dates}
\tablehead{\colhead{Field} & \colhead{Filter} & \colhead{Observation Date} & \colhead{FWHM PSF} & \colhead{$\alpha$(J2000)} & \colhead{$\delta$(J2000)}} 
\startdata
SFF01  & r & 09/17/2017 & 0.89\arcsec &21:42:42 &19:59:28 \\
 & i & 09/17/2017 & 0.83\arcsec  &&\\
 & g & 09/17/2017 & 0.76\arcsec && \\
 & NB1 & 09/13/2018 & 0.93\arcsec  &&\\
 & NB2 & 09/14/2018 & 0.78\arcsec  &&\\
 & NB3 & 09/13/2018 & 0.96\arcsec  &&\\
 \hline
SFF10  & r & 11/07/2016 & 0.83\arcsec  & 01:44:20 &27:54:13 \\
 & i & 08/19/2017 & 0.81\arcsec && \\
 & g & 08/19/2017 & 1.23\arcsec  &&\\
 & NB1 & 11/07/2016 &  0.85\arcsec  &&\\
 & NB2 & 09/14/2018 &  0.85\arcsec &&\\
 & NB3 & 09/13/2018 &  0.69\arcsec  &&\\
 \hline
SFF15  & r & 11/07/2016 & 0.81\arcsec & 02:38:52 &27:51:43\\
 & i & 08/19/2017 & 0.87\arcsec &&\\
 & g & 08/19/2017 & 1.17\arcsec  &&\\
 & NB1 & 11/07/2016 & 0.72\arcsec &&\\
 & NB2 & 09/14/2018 & 0.70\arcsec &&\\
 & NB3 & 09/14/2018 & 0.66\arcsec &&\\
\enddata
\end{deluxetable}

\begin{figure}\label{fig:filters}
    \centering
    \includegraphics[scale=0.5]{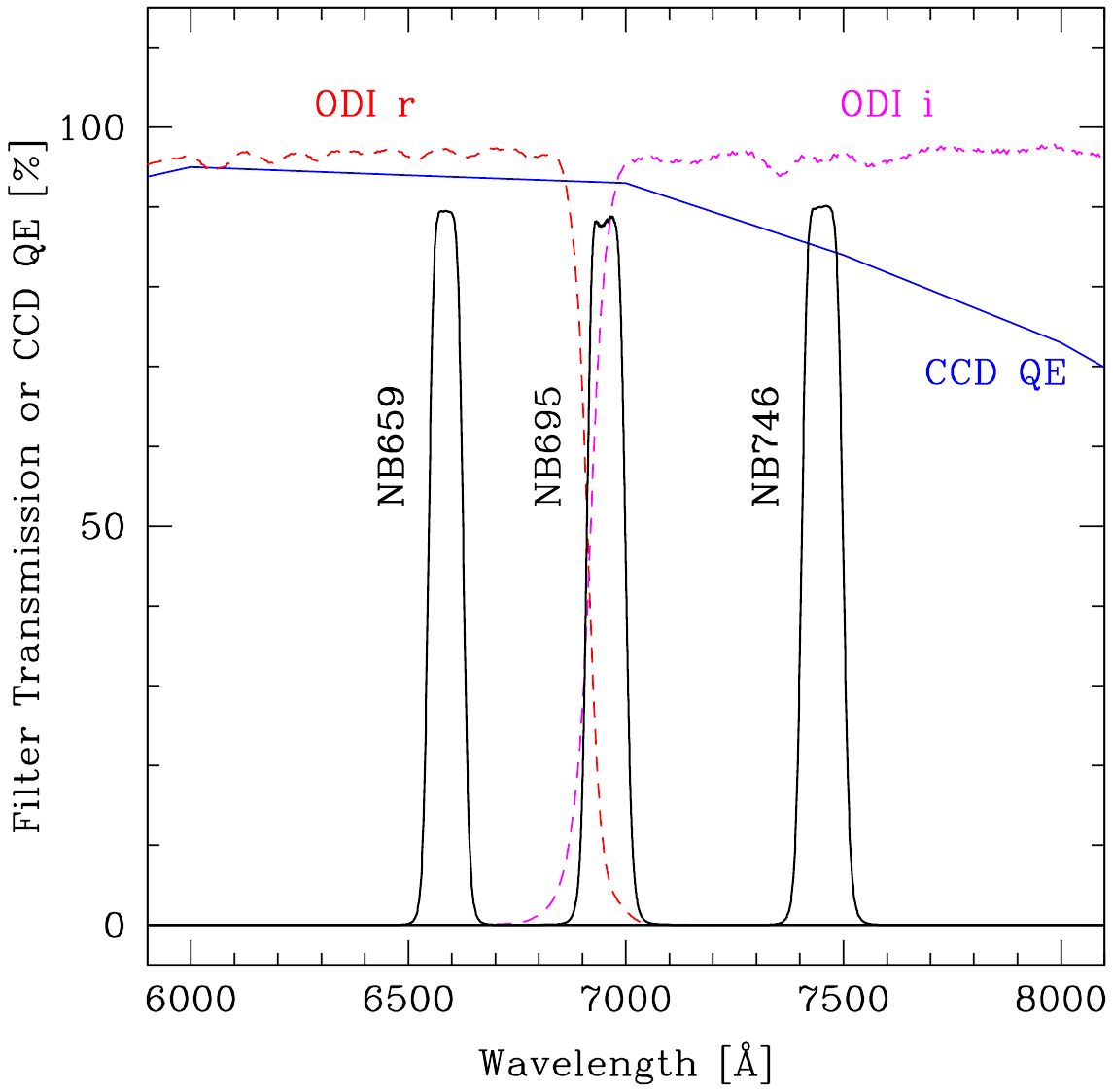}
    \caption{The filter transmission curves of our three narrow-band filters: NB659 (NB2), NB695 (NB1), and NB746 (NB3). The dashed lines show part of the transmission curves for the r and i broad-band filters. Overlaid is the quantum efficiency (QE) curve of the CCD (solid line), demonstrating that while it does start to drop off around 7000~\AA, the sensitivity is still high in i and NB3.}
    \label{fig:filters}
\end{figure}

When using ODI, telescope tracking occurs using a star located on one of the OTA chips which is read out continuously during the exposure at video rates.  Because the OTA chip used for guiding is lost to the science image, we attempted to select guide stars located on different OTAs for each image in the dither sequence.  This avoids large unusable areas in our final stacked images. 


\subsubsection{Calibration Observations}
Following standard observing procedures, bias and dark images were taken each night. This included 10 zero-second bias frames followed by three 600-second dark current images. These are crucial for correcting detector signatures during the initial processing. Spectrophotometric standard star observations were also taken; these are further discussed in section~\ref{sec:calibration_process}.

Flat field images are taken by the WIYN staff approximately once per month through each filter and are applied to the processing of the recently-taken data. A special technique is employed. A slow shutter blade speed is used in order to baffle out internal reflections, and thus eliminate the pupil ghost. The slow shutter technique works such that both shutters move at once, with only a small delay between them, effectively creating a slit aperture which moves across the frame. Raw flat field images are acquired with at least two different rotations of the instrument so that any gradients because of non-uniform illumination of the flat field can be smoothed out. The stability of the flats is very good, with variations of less than 1\% over many months.

\subsection{Data Processing}\label{ssec:proc}
Raw images are processed and analyzed utilizing both the ODI Pipeline, Portal, and Archive (PPA), as well as custom scripts written in the Image Reduction and Analysis Facility ({\tt IRAF}) and {\tt Python}. Each of the processing steps are detailed in the following sub-sections.

\subsubsection{ODI Pipeline, Portal, and Archive}

Raw ODI images are transferred from WIYN to the PPA, which is hosted by Indiana University.  In the PPA, the raw data are first run through the \texttt{QuickReduce} pipeline~\citep{QUICKREDUCE}, which begins by masking out unusable pixels, whether this is due to persistency, trailing, a defective cell, cross-talk, or a static bad pixel. Overscan levels, bias, and dark levels are determined and subtracted from each of the raw images. A correction based on the flat fields and any known non-linearity between the observed counts and the exposure is then applied. The final step is an astrometric calibration which is performed using Gaia~\citep{GaiaDR1} as the reference catalog. The output from \texttt{QuickReduce} is one complete FITS image for each dither position, properly reduced and ready for further processing.

Next, the astrometric mapping software {\tt SWarp}~\citep{SWarp} is run from within the PPA, which aligns and combines all of our images from each dither sequence to produce one image for each filter for each field. We use a weighted combination mode, an illumination correction, and utilize a surface fit for the background subtraction method, preserving extended objects of at least 3\arcmin. The process also masks bad pixels and removes the OTAs used in guiding. The output image is re-projected with a new pixel scale of 0.125\arcsec\ pixel$^{-1}$.

\subsubsection{SFACT Pre-processing Steps}\label{ssec:preproc}
The reduced and stacked ODI images are retrieved from the PPA for subsequent processing. The image from each filter is cropped such that all images for each field cover exactly the same area on the sky and are precisely aligned with one another. This ensures that the objects identified in a field have the same positions in each filter later in the processing. 

A master image is then created by summing all six individual images together, resulting in a very deep image. Objects as faint as r $\sim$ 26 are readily detected in the master image. This deep master image is used for catalog creation as discussed in Section~\ref{Sec:Target selection}. Because this image includes both narrow- and broad-band filters, it allows for the detection of ELGs which have extremely faint continuum flux but strong nebular emission, which would otherwise be missed in a BB-only image.

The average point spread function (PSF) full-width at half-maximum (FWHM) is determined using roughly a dozen user-selected stars. This measure of the image quality is determined for the master image as well as the individual filter images. All images are then binned 2$\times$2, resulting in a final image scale of 0.25\arcsec\ pixel$^{-1}$.  This value was chosen because a native resolution (seeing) better than 0.5\arcsec\ pixel$^{-1}$ is only rarely obtained at WIYN.  While our objects tend to have small angular sizes, they are almost never undersampled with this choice of pixel scale.  Finally,  a script is run on the binned master image which allows the user to select an object-free region in order to determine the background noise level, a crucial parameter used during the object detection stage. 

The final step in the pre-processing of the images is to create scaled {\it continuum images} appropriate for each NB image.   These scaled images, which are derived from our survey BB images, are used as part of our object-selection process (described in Section \ref{Sec:Target selection}), as well as for creating continuum-free difference images (as illustrated in Section 4.2).  

As seen in Figure~\ref{fig:filters}, the NB1 filter is located at the transition between the r-band and i-band filters, while NB2 is located in the redder half of the r-band filter.  As a result of these locations, extremely red objects (e.g., M stars and high-redshift early-type galaxies) show a significant flux excess in both NB1 and NB2 when only the r-filter image is used as the continuum image, leading to false detections.  Tests revealed that the sum of the r- and i-filter images provides far better results as the continuum images for both NB1 and NB2 than does using the r-band image alone.   On the other hand, the i-band image proves adequate for use as the continuum image for NB3.  

As a result of our evaluation we adopt, the sum of the r-band and i-band images (r + i) as the continuum image for the NB1 and NB2 images and the i-band image as the continuum image for NB3.  All three continuum images are then scaled to match the flux levels in the individual NB filter images.  This is done by measuring the fluxes of approximately a dozen user-selected stars in both the continuum and NB images and then scaling the continuum image to match the flux in the NB image.  These scale factors account for numerous factors, such as the differences in filter bandwidths, exposure times, airmass and sky transparency.  The first two terms always dominate, meaning that the scale factors derived this way have characteristic values that reflect the ratios of the filter bandwidths and exposure times.  For example, the ratio of the bandwidths of the r + i and NB1 filters is $\sim$28.6, while the ratio of the exposure times is 0.2.  This results in an expected scale factor of $\sim$5.7, which is the middle of the range of measured scale factors we determine.   Similarly, the scale factors for the NB2 and NB3 continuum images have characteristic values of $\sim$6.4 and $\sim$2.7.

\section{SFACT Object Selection and Photometry}\label{Sec:Target selection}

In the following section we detail the methods carried out to select and measure emission-line candidates from our survey images.   The first step utilizes the ultra-deep master image of each field to create a comprehensive catalog of all objects detected within this image.  Next we perform small-aperture photometry on every object in the catalog using the continuum and NB images and identify those sources that exhibit an excess of flux in the NB image, indicating a potential ELG or QSO.  All candidates are then checked visually to remove objects that are image artifacts.   Once the final list of SFACT candidates is established, each source in the final catalog of ELGs is carefully measured in all three broad-band images ({\it gri}) as well as the relevant NB image.

Each survey field typically contains on the order of 10$^5$ total objects detected at the sensitivity limit of our master image. Custom scripts were written to identify relevant objects in a fully automated process. These scripts were implemented to reduce the large number of objects that needed to be evaluated as possible ELG candidates in each field to a manageable level. Manual verification was performed as a last step.  We perform the following analysis on quadrants (designated A-D) of our full-frame images in order create more manageable data sets for the user. The quadrants were created with 100 pixel overlaps to ensure that objects were not missed along boundaries. 

\subsection{Master Catalogs of Sources}\label{ssec:catalog}

The first phase of our analysis focuses on creating a list of all objects detected in the six-filter summed master images.  The characteristic limiting magnitudes of our master images are r $\sim$ 25.5-26.0.  The combined six-filter master images yields a greater number of faint objects than would be possible from the individual filter images.   Additionally, since one of the goals of SFACT is to catalog all emission-line sources in each field, our catalog method needs to be sensitive to both nearby bright, extended sources and more distant faint, unresolved objects (and everything in between).  

We utilize {\tt DAOFIND}~\citep{DAOPHOT} for the purpose of automatically detecting every object within our survey fields.   The searches are carried out by running {\tt DAOFIND} multiple times using a series of image kernels of various sizes in order to detect objects with a range of light distributions.  This allows for the identification of small compact objects as well as larger, extended galaxies.   Since the multiple runs of {\tt DAOFIND} detect most objects multiple times, we scan the final catalog and remove all duplicate entries before proceeding.

Next we convert the image positions (x,y) to sky coordinates (RA, Dec), after which we carry out an identification of cross-matches within the SDSS database. While many of our objects are too faint to be identified in SDSS, for those that are we collect additional information to add to our database, including photometry and SDSS classification (star or galaxy).   After this automatic processing, we visually check the master image for bad regions to mask. This step is carried out to mitigate problems with sections of the image which, due to several non-functioning OTA cells, do not yield usable data. We mark these regions as a series of boxes and any object within this region is removed from future consideration. The area contained within these masked regions is recorded in the header of the catalog table and removed from the total area of the field when doing computations involving the survey area.

It is worth stressing that our master catalog of sources can be applied to each of the individual filter images since each of these images was carefully co-aligned prior to making the six-filter master image.  That is, any source in the master catalog will have the same position in each of the individual filter images.  Hence, this deep master catalog serves as the basis for all subsequent searches for ELGs in the individual NB images.

\subsection{Identifying SFACT Candidates}\label{ssec:indentifying}

Once the object cataloging phase is completed we measure the instrumental magnitudes for every object in each NB image plus continuum image pair.  We note that the NB images are not continuum subtracted and, because of the flux scaling described above, a source with no emission should have the same instrumental magnitude is both images.  

The instrumental magnitudes are measured in small apertures that are set to a diameter of 3 times the FWHM of the stellar PSF relevant for each image.   Since all objects in the master catalog are measured, our procedure measures isolated point sources as well as knots of emission in extended galaxies.  That is, this methodology is sensitive to detecting emission-line objects regardless of their sizes or morphologies as long as the objects has been identified in our cataloging process described above.   We designate these instrumental magnitudes as $m_{NB}$  for measurements of the NB images and $m_{cont}$ for measurements of the continuum images.

Based on these instrumental magnitudes, we next perform a secondary scaling step as a fine-tuning offset calculation.  While the NB and continuum images have already been scaled to each other (see  Section~\ref{ssec:preproc}), it was found that this preliminary scaling based on typically 10-15 stars was not always accurate.   Hence, a secondary scaling using our photometry for many dozens of stars was carried out.
All stars which have $m_{cont} < -10.5$ are used to compute an offset such that the median $\Delta m$ = 0 for these stars.  Using the median of these offset values, a quadrant-wide offset is determined and applied to all of the objects in the table.    This scaling offset is typically small, ranging between 0.00 and 0.15 magnitudes.   

\begin{figure}[t]
    \centering
    \includegraphics[scale=1]{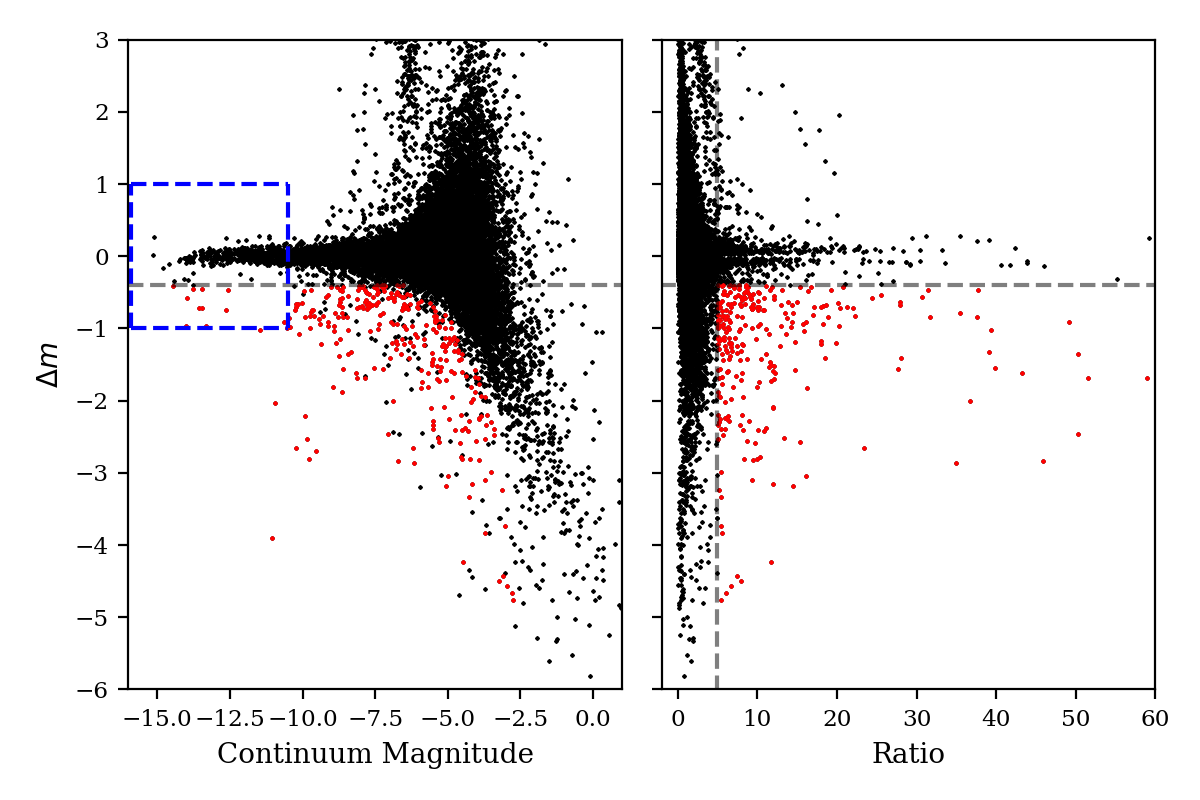}
    \caption{An example diagnostic plot for the SFF01 field. On the left we plot $\Delta m$ against the continuum magnitude. On the right, we plot the $\Delta m$ against the {\it ratio}. The horizontal line represents our $\Delta m$ cutoff of -0.4; the vertical line represents our {\it ratio} cutoff of 5.0. Inside the blue box in the left panel are the stars used to refine the $\Delta m$ offsets. All objects within the lower right section of the right plot are considered possible candidates, subject to further analysis.  These objects are shown in red in both plots.}
    \label{fig:diag}
\end{figure}

Using the instrumental magnitudes, we compute the magnitude difference ($\Delta m$) for each source in the catalog:
\begin{equation}\label{eq:magdiff}
\Delta m=m_{NB}-m_{cont}.
\end{equation}
We plot $\Delta m$ versus m$_{cont}$ for the objects in SFF01 in the lefthand plot of Figure~\ref{fig:diag}.   The blue dashed line designates the stars used to compute the secondary offset correction.   This correction helps ensure that all of the continuum flux is properly removed in the calculation of $\Delta m$.  We also measure a pseudo signal-to-noise ratio (henceforth referred to simply as {\it ratio}) for each object. We use
\begin{equation}\label{eq:snr1}
\sigma_{\Delta m} = ( \sigma_{NB}^2 + \sigma_{cont}^2 )^{\frac{1}{2}}
\end{equation}
\begin{equation}\label{eq:snr2}
ratio=\frac{\Delta m}{\sigma_{\Delta m}}
\end{equation}
where $\sigma_{NB}$ is the uncertainty in $m_{NB}$ and $\sigma_{cont}$ is the uncertainty in $m_{cont}$. 
The righthand portion of Figure~\ref{fig:diag} plots $\Delta m$ versus {\it ratio} for the objects in the SFF01 field.

We use $\Delta m$ and {\it ratio} to indicate which objects have a statistically significant excess of flux in the NB filter. 
That is, objects with a large negative value of $\Delta m$ have significantly more flux in the NB image than in the continuum image, while objects with larger values of {\it ratio} are statistically more significant.
We experimented with a range of values for $\Delta m$ and {\it ratio} to be used for our ELG selection criteria, running tests on multiple fields before selecting our final limits.
In addition, we used our experience with previous NB emission-line surveys \citep[e.g.,][]{hadot1, KISS1} as a guide for reasonable values for the limits.  
We settled upon using values of $\Delta m$ lower than -0.4 and {\it ratio} greater than 5.0 for inclusion in our ELG candidate list as providing the best balance between the desire to select candidate objects which are as faint as possible, while minimizing the number of false detections.
This region of parameter space is delimited in the righthand plot of Figure~\ref{fig:diag} by horizontal and vertical dashed lines.  Objects in the lower right section of this plot (shown in red) are ELG candidates.

The quantity $\Delta m$ can be related directly to the equivalent width of the detected emission lines by simply applying the definition of a magnitude:
\begin{equation}\label{eq:ew}
\Delta m = m_{NB}-m_{cont} = -2.5\cdot log\left( \frac{f_{NB}}{f_{cont}}\right) \approx -2.5\cdot log\left( \frac{f_{line}+f_{cont}}{f_{cont}}\right) = -2.5\cdot log\left( \frac{f_{line}}{f_{cont}}+1\right)
\end{equation}
Our $\Delta m$ selection threshold of $-$0.4 mag thus corresponds to a flux ratio of f$_{line}$/f$_{cont}$ $\approx$ 0.445.   With our NB filter widths of 81 to 97~\AA\ (see SFACT1), the $\Delta m$ limit corresponds to rest-frame equivalent width selection limits of $\sim$36-39~\AA\ for \HA detections, $\sim$27-30~\AA\ for [\ion{O}{3}] detections, $\sim$20-22~\AA\ for [\ion{O}{2}] detections.

The final stage in the candidate-selection process involves checking and verifying the objects identified via their $\Delta m$ and {\it ratio} values.  Additional scripts are used to automatically filter out obvious false detections, including saturated pixels and bleed trails associated with bright stars, image artifacts such as bad pixels, and strong cosmic rays.  Once the software has selected a cleaned list of possible candidates, members of our team evaluate image cutouts created for each object (similar to what is shown in Figures~\ref{Fig:HA detections} --~\ref{find:C21205}).  This visual inspection is carried to out to ensure that the selected objects are real (i.e., not image artifacts) and that they possess a detectable amount of NB flux above the level seen in the continuum image.  Each team member makes an independent decision on whether or not each object is a valid ELG candidate, and any object that is questionable is vetted by the group before a decision is made on whether or not to accept or reject it.  Typically there are of order 100-200 objects per field per filter identified as SFACT candidates by our automated software, but approximately 50-75\% are rejected as spurious during this manual checking process. Most of these are image artifacts. 

Because of the way the SFACT survey is carried out, we often detect multiple \HII regions in a single spiral galaxy.  This applies only to the relatively low-redshift H$\alpha$-detected objects (z $<$ 0.15).  The SFACT program is primarily concerned with the global properties of each emission-line galaxy, rather than carrying out an accounting of each  \HII region.  Therefore, we do not retain every single \HII region in our final catalog of sources.  As discussed in SFACT1, we retain only the most prominent \HII region as a target for spectroscopic follow-up, and we use the center of the galaxy for carrying out global photometry.  The lists of the individual \HII regions are retained and may become part of a separate study in the future.

Following the procedures specified in this section, we arrived at final catalogs of ELG candidates for the three pilot-study fields.  We detected 132 SFACT objects in SFF01, 216 in SFF10, and 185 in SFF15.  All of these objects are then slated for spectroscopic observation, as discussed in SFACT3.

\vfill\eject
\subsection{Photometry}\label{sec:photometry}

After the object selection is complete we perform photometry of the SFACT candidates in the three BB ({\it gri}) images, as well as the relevant continuum-subtracted NB image.
Here we discuss our method for performing the calibration of the BB instrumental magnitudes, the two-step process for correctly putting the NB fluxes on an appropriate flux scale, and our procedure for determining accurate brightness measurements for all SFACT targets. 

\subsubsection{Calibration}\label{sec:calibration_process}
All SFACT imaging products are calibrated utilizing photometric information from SDSS stars present in our science images.  For this purpose we limit the use of SDSS stars to those with r-band magnitudes brighter than 20.0 and a g-r color in the range 0.4 -- 1.1.  We further restrict the sample of SDSS stars in each filter by an upper limit on the photometric error, typically between 0.02 and 0.03 magnitude.
There exist sufficient numbers of stars in each field that satisfy these criteria to ensure robust photometric calibrations.   We perform aperture photometry on each SDSS star and compare the derived instrumental magnitudes with the tabulated SDSS photometry, providing a difference value for each calibration star ($\Delta$m(SDSS)).  We compute the mean and standard deviation of the $\Delta$m(SDSS) values for all stars in each filter, retaining those within $3\sigma$, iterating this once to remove outliers.  This cleaned sample contains hundreds of stars in each BB and NB filter. A final mean $\Delta$m(SDSS) for each filters is calculated for this clean sample and used as the zero point constant (ZPC) of the entire field (ZPC(SDSS)).

In order to evaluate the homogeneity of the SDSS-based calibration across our spatially-large images, we divide each field into nine sections. Each section of the image typically has several dozen stars per filter. In each section, we compute the mean difference of the stars: a section-specific ZPC(SDSS). In our three pilot fields, no significant positional differences in the ZPC values across the images have been found.   Most variations between the nine sections are less than $\pm$0.01 mag.

We also utilize the section-specific ZPC(SDSS) values to derive an estimate of the uncertainty of the overall ZPC(SDSS) for each field. This is done because computing a formal uncertainty in the main ZPC by adopting a more traditional $\sigma/\sqrt{N}$ type error in the mean results in an unphysically small ZPC uncertainty due to the large number of SDSS stars in each field. As an alternative, we determine the standard deviation of the means in each of the nine sections. This standard deviation is used as our estimate of the uncertainty in the ZPC(SDSS) of the entire field for each filter.   Characteristic uncertainties range between 0.005 and 0.015 mag for both the BB and NB filters.

Due to the narrow filter bandwidth, NB photometric calibrations traditionally do not use a color term \citep[e.g.,][]{LVLHA}, a convention followed by this study. For BB photometry, we expected the color terms to be extremely small since the filters utilized are similar to those used by SDSS. We used our measurements from the pilot-study fields to verify that the color terms for the r and i filters are vanishingly small. The color term for g is somewhat larger ($\epsilon_g=0.105 \pm 0.002$), and we have applied it to our g-band magnitudes.

\subsubsection{NB Offset Calibration}\label{ssec:NBoffset}
The calibration of our NB flux measurements requires an additional step. This step utilizes observations of spectrophotometric standard stars (e.g.,~\citealp{Oke, Massey}) that were observed through each of our NB filters.   In the NB SFACT science images, we perform the initial calibration using the SDSS stars just like what is performed on the BB images. This produces a magnitude difference between our SFACT objects and the SDSS stars. Because the instrumental magnitudes are measured in the same image, time-dependent quantities such as atmospheric extinction are effectively accounted for. 

To properly place our NB measurements on an appropriate flux scale, we then perform an additional offset calibration utilizing observations of spectrophotometric standard stars. We repeat the same measurement procedure described above using the spectrophotometric standard stars as the ``science target" and the SDSS stars found in the standard star images as the calibration sources. We arrive at a magnitude difference between the SDSS stars and our standard star. Because we employ the same filters for the science images and the calibration images, the ZPC of the SDSS stars will be the same. We can make use of this equivalence to place our NB measurements on an absolute flux scale.  The offset calibration derived in this way ranges from 0.13 to 0.33 mag for the three NB filters.  This offset calibration is applied on a filter by filter basis to all SFACT objects to complete the NB calibration.

\subsubsection{Aperture Photometry}\label{ssec:apphot}
We perform photometric measurements on each SFACT object using a range of apertures. The aperture-determination step is performed on the master image in order to determine the proper BB photometric aperture for each object. We carry out a curve-of-growth analysis to determine the optimal aperture to use.
Photometry is then performed on each of the individual filter images using the appropriate aperture. 

Because many \HII regions are located in extended galaxies, appearing as multiple knots of emission, determining the correct aperture to use is a challenge. Moreover, light from the rest of the galaxy will always be conflated with the light from the \HII region. For the sake of uniformity, all \HII regions are assigned the same aperture of 16 pixels (4\arcsec). This size has been chosen through trial and error since it adequately encapsulates the light from each individual \HII region.   While we measure and record the photometric properties of the individual \HII regions, for most applications we utilize the measurements obtained for the entire galaxy.

If the curve-of-growth analysis in our script does not converge on an aperture to use, we examine the object by eye. We display a tiled image of each BB filter image as well as the master image overlaid with the suggested aperture. We use an interactive process to manually select an aperture which best captures the light from the target. We also inspect and confirm the apertures of sources which have near neighbors to avoid possible contamination of the photometry from the nearby source.  Once we have instrumental magnitudes in all of the BB filters, we apply the ZPCs previously calculated which puts our objects on the same magnitude scale as SDSS. 

Based on the results of our BB photometry, we are able to establish the depths of our individual BB images.   Characteristic 3$\sigma$ limiting magnitudes for the SFACT images are 24.4 mag in the r-band, 24.0 in the i-band, and 25.5 in the g-band.

The photometric measurement process used for the NB images follows a procedure similar to the one described above for the BB data, with only minor differences. The curve-of-growth analysis is performed only on the continuum-subtracted NB image corresponding to the NB filter the ELG was detected in. This ensures that we are determining an aperture based on an image which contains only the emission-line flux we want to measure.  Once again, we visually check the aperture of any target where the curve-of-growth software does not yield a robust solution.  Once the final NB instrumental magnitudes are measured, we apply the NB ZPC calculated in the same way as the BB ZPC. We also apply the NB calibration offset described in Section~\ref{ssec:NBoffset} to place our NB measurements on a proper flux scale.  Our NB magnitudes are then converted into line fluxes using the calibration from \citet{Massey}:  m$_\nu$ = $-$2.5 $\cdot$ log(f$_\nu$) $-$ 48.29.



\begin{deluxetable*}{ccccccccccc}
    \tabletypesize{\scriptsize}
\rotate
\caption{SFF01 Emission-Line Objects \label{tab:SFF01}}
    \tablehead{
    \\
    SFACT Object ID & SFACT Coordinate ID & $\alpha$(J2000) & $\delta$(J2000) & $\Delta$m & ratio & Object & m$_r$ & m$_i$ & m$_g$ & log(f$_{NB}$)\\
    & & H:M:S & D:M:S & mag & & Type & mag & mag & mag &  erg/s/cm$^2$ \\  
    (1) & (2) & (3) & (4) & (5) & (6) & (7) & (8) & (9) & (10) & (11)
    }
 \startdata
 SFF01-NB3-D20110 & SFACT J214123.25+200510.7 &   21:41:23.25 &   20:05:10.7 &  -0.57 $\pm$   0.02 &  30.80 & ExtG &  18.378 $\pm$  0.011 &  18.031 $\pm$  0.009 &  18.972 $\pm$  0.009 & -14.028 $\pm$   0.009 \\
 SFF01-NB3-D20084 & SFACT J214123.34+200509.5 &   21:41:23.34 &   20:05:09.5 &  -0.93 $\pm$   0.02 &  45.73 & HII &  19.370 $\pm$  0.012 &  19.030 $\pm$  0.011 &  19.841 $\pm$  0.011 & -14.262 $\pm$   0.009 \\
 SFF01-NB3-D19969 & SFACT J214123.61+201118.8 &   21:41:23.61 &   20:11:18.8 &  -0.64 $\pm$   0.10 &   6.35 & ELG &  21.733 $\pm$  0.057 &  21.706 $\pm$  0.089 &  22.537 $\pm$  0.064 & -15.484 $\pm$   0.076 \\
 SFF01-NB3-B20552 & SFACT J214126.31+195845.4 &   21:41:26.31 &   19:58:45.4 &  -0.37 $\pm$   0.01 &  25.23 & ExtG &  18.323 $\pm$  0.010 &  17.868 $\pm$  0.007 &  19.101 $\pm$  0.008 & -14.395 $\pm$   0.012 \\
 SFF01-NB1-B20542 & SFACT J214126.46+194344.1 &   21:41:26.46 &   19:43:44.1 &  -0.76 $\pm$   0.12 &   6.47 & ELG &  22.904 $\pm$  0.072 &  22.683 $\pm$  0.098 &  23.567 $\pm$  0.087 & -15.667 $\pm$   0.056 \\
 SFF01-NB2-D19115 & SFACT J214126.46+201342.6 &   21:41:26.46 &   20:13:42.6 &  -0.47 $\pm$   0.03 &  14.98 & ELG &  20.313 $\pm$  0.016 &  19.781 $\pm$  0.016 &  21.779 $\pm$  0.028 & -14.965 $\pm$   0.018 \\
 SFF01-NB3-B20497 & SFACT J214126.52+195850.3 &   21:41:26.52 &   19:58:50.3 &  -0.49 $\pm$   0.02 &  20.91 & ExtG &  19.083 $\pm$  0.011 &  18.773 $\pm$  0.011 &  19.553 $\pm$  0.010 & -14.484 $\pm$   0.016 \\
 SFF01-NB3-B19399 & SFACT J214131.74+195847.3 &   21:41:31.74 &   19:58:47.3 &  -0.64 $\pm$   0.02 &  30.05 & ELG &  19.215 $\pm$  0.013 &  18.858 $\pm$  0.013 &  19.701 $\pm$  0.012 & -14.376 $\pm$   0.012 \\
 SFF01-NB2-D17902 & SFACT J214132.16+202146.6 &   21:41:32.16 &   20:21:46.6 &  -0.70 $\pm$   0.09 &   7.96 & ELG &  22.346 $\pm$  0.093 &  21.566 $\pm$  0.071 &  22.880 $\pm$  0.095 & -15.285 $\pm$   0.045 \\
 SFF01-NB2-B19207 & SFACT J214132.90+193945.5 &   21:41:32.90 &   19:39:45.5 &   0.17 $\pm$   0.05 &   3.34 & ExtG &  18.747 $\pm$  0.014 &  18.626 $\pm$  0.017 &  18.951 $\pm$  0.010 & -14.649 $\pm$   0.022 \\
 \\
 SFF01-NB2-B19198 & SFACT J214132.93+193942.1 &   21:41:32.93 &   19:39:42.1 &  -0.61 $\pm$   0.07 &   8.14 & HII &  21.146 $\pm$  0.036 &  21.086 $\pm$  0.044 &  21.451 $\pm$  0.033 & -15.377 $\pm$   0.039 \\
 SFF01-NB1-B19076 & SFACT J214133.33+194226.7 &   21:41:33.33 &   19:42:26.7 &  -0.82 $\pm$   0.14 &   6.01 & ELG &  23.427 $\pm$  0.119 &  22.639 $\pm$  0.096 &  23.351 $\pm$  0.071 & -15.726 $\pm$   0.069 \\
 SFF01-NB3-B18885 & SFACT J214134.29+194341.1 &   21:41:34.29 &   19:43:41.1 &  -0.97 $\pm$   0.10 &   9.97 & ELG &  22.740 $\pm$  0.058 &  22.380 $\pm$  0.071 &  22.924 $\pm$  0.053 & -15.419 $\pm$   0.035 \\
 SFF01-NB3-B18561 & SFACT J214135.87+194757.5 &   21:41:35.87 &   19:47:57.5 &  -0.69 $\pm$   0.13 &   5.12 & ELG &  23.252 $\pm$  0.101 &  22.674 $\pm$  0.109 &  23.708 $\pm$  0.105 & -15.776 $\pm$   0.062 \\
 SFF01-NB2-B18506 & SFACT J214136.06+195653.7 &   21:41:36.06 &   19:56:53.7 &  -0.81 $\pm$   0.15 &   5.52 & ELG &  23.401 $\pm$  0.108 &  23.098 $\pm$  0.139 &  23.873 $\pm$  0.109 & -15.804 $\pm$   0.050 \\
 SFF01-NB3-B18371 & SFACT J214136.92+193734.2 &   21:41:36.92 &   19:37:34.2 &  -1.30 $\pm$   0.17 &   7.60 & ELG &  23.484 $\pm$  0.119 &  23.154 $\pm$  0.140 &  23.727 $\pm$  0.097 & -15.574 $\pm$   0.047 \\
 SFF01-NB3-B17245 & SFACT J214141.87+195707.7 &   21:41:41.87 &   19:57:07.7 &  -1.02 $\pm$   0.14 &   7.27 & ELG &  22.817 $\pm$  0.073 &  22.685 $\pm$  0.108 &  22.971 $\pm$  0.059 & -15.452 $\pm$   0.051 \\
 SFF01-NB3-D15415 & SFACT J214142.73+201252.3 &   21:41:42.73 &   20:12:52.3 &  -0.68 $\pm$   0.10 &   6.89 & ELG &  22.246 $\pm$  0.055 &  22.249 $\pm$  0.090 &  22.786 $\pm$  0.067 & -15.581 $\pm$   0.058 \\
 SFF01-NB2-D15191 & SFACT J214143.48+200448.4 &   21:41:43.48 &   20:04:48.4 &  -0.61 $\pm$   0.08 &   7.47 & ELG &  21.655 $\pm$  0.049 &  21.398 $\pm$  0.062 &  22.484 $\pm$  0.059 & -15.466 $\pm$   0.046 \\
 SFF01-NB1-B16317 & SFACT J214146.59+194328.3 &   21:41:46.59 &   19:43:28.3 &  -0.75 $\pm$   0.11 &   6.97 & ELG &  22.463 $\pm$  0.086 &  22.048 $\pm$  0.097 &  22.913 $\pm$  0.092 & -15.647 $\pm$   0.056 \\
\\
 SFF01-NB3-B16011 & SFACT J214148.07+195758.9 &   21:41:48.07 &   19:57:58.9 &  -1.10 $\pm$   0.11 &  10.46 & ELG &  22.690 $\pm$  0.056 &  22.609 $\pm$  0.087 &  23.057 $\pm$  0.057 & -15.367 $\pm$   0.038 \\
 SFF01-NB3-B15732 & SFACT J214149.54+194038.6 &   21:41:49.54 &   19:40:38.6 &  -0.67 $\pm$   0.10 &   6.99 & ELG &  22.540 $\pm$  0.070 &  22.134 $\pm$  0.074 &  22.934 $\pm$  0.061 & -15.548 $\pm$   0.065 \\
 SFF01-NB2-B15722 & SFACT J214149.60+193848.3 &   21:41:49.60 &   19:38:48.3 &  -0.42 $\pm$   0.07 &   5.56 & ELG &  22.185 $\pm$  0.058 &  21.341 $\pm$  0.044 &  23.128 $\pm$  0.077 & -15.670 $\pm$   0.054 \\
 SFF01-NB1-D13076 & SFACT J214150.05+200728.2 &   21:41:50.05 &   20:07:28.2 &  -0.96 $\pm$   0.13 &   7.26 & ELG &  23.247 $\pm$  0.104 &  22.684 $\pm$  0.110 &  23.800 $\pm$  0.113 & -15.705 $\pm$   0.053 \\
 SFF01-NB1-D12776 & SFACT J214150.91+202341.9 &   21:41:50.91 &   20:23:41.9 &  -0.51 $\pm$   0.08 &   6.04 & ELG &  22.214 $\pm$  0.085 &  21.255 $\pm$  0.053 &  23.238 $\pm$  0.114 & -15.577 $\pm$   0.057 \\
 SFF01-NB3-B14965 & SFACT J214153.12+195235.0 &   21:41:53.12 &   19:52:35.0 &  -0.42 $\pm$   0.03 &  14.55 & ELG &  19.891 $\pm$  0.016 &  19.600 $\pm$  0.019 &  20.436 $\pm$  0.016 & -14.907 $\pm$   0.021 \\
 SFF01-NB3-B14799 & SFACT J214153.89+195227.9 &   21:41:53.89 &   19:52:27.9 &  -0.62 $\pm$   0.11 &   5.74 & ELG &  22.701 $\pm$  0.072 &  22.138 $\pm$  0.075 &  22.944 $\pm$  0.066 & -15.595 $\pm$   0.059 \\
 SFF01-NB3-D11444 & SFACT J214155.02+201511.4 &   21:41:55.02 &   20:15:11.4 &  -0.65 $\pm$   0.09 &   7.22 & ELG &  22.600 $\pm$  0.050 &  22.323 $\pm$  0.063 &  23.024 $\pm$  0.060 & -15.619 $\pm$   0.047 \\
 SFF01-NB2-B14205 & SFACT J214156.04+195310.0 &   21:41:56.04 &   19:53:10.0 &  -0.83 $\pm$   0.08 &  10.89 & ELG &  21.993 $\pm$  0.048 &  21.888 $\pm$  0.069 &  22.616 $\pm$  0.054 & -15.308 $\pm$   0.031 \\
 SFF01-NB1-D10796 & SFACT J214157.21+201512.0 &   21:41:57.21 &   20:15:12.0 &  -0.92 $\pm$   0.10 &   9.25 & ELG &  22.868 $\pm$  0.068 &  22.466 $\pm$  0.073 &  25.806 $\pm$  0.509 & -15.471 $\pm$   0.044 \\
  \enddata

\tablecomments{Table \ref{tab:SFF01} is published in its entirety in the machine-readable format. A portion is shown here for guidance regarding its form and content.}
\end{deluxetable*}



\begin{deluxetable*}{ccccccccccc}
    \tabletypesize{\scriptsize}
\rotate
\caption{SFF10 Emission-Line Objects \label{tab:SFF10}}
    \tablehead{
    \\
    SFACT Object ID & SFACT Coordinate ID & $\alpha$(J2000) & $\delta$(J2000) & $\Delta$m & ratio & Object & m$_r$ & m$_i$ & m$_g$ & log(f$_{NB}$)\\
    & & H:M:S & D:M:S & mag & & Type & mag & mag & mag &  erg/s/cm$^2$ \\  
    (1) & (2) & (3) & (4) & (5) & (6) & (7) & (8) & (9) & (10) & (11)
    }
 \startdata
 SFF10-NB3-D13755 & SFACT J014256.70+281615.3 &    1:42:56.70 &   28:16:15.3 &  -1.72 $\pm$   0.28 &   6.26 & ELG &  25.846 $\pm$  0.986 &  24.064 $\pm$  0.189 &  25.491 $\pm$  0.370 & -15.572 $\pm$   0.053 \\
 SFF10-NB3-D13569 & SFACT J014258.14+275740.4 &    1:42:58.14 &   27:57:40.4 &  -0.72 $\pm$   0.08 &   8.46 & ELG &  22.157 $\pm$  0.087 &  22.058 $\pm$  0.073 &  23.037 $\pm$  0.074 & -15.299 $\pm$   0.038 \\
 SFF10-NB2-B12883 & SFACT J014258.90+274309.1 &    1:42:58.90 &   27:43:09.1 &  -1.12 $\pm$   0.17 &   6.55 & ELG &  23.580 $\pm$  0.167 &  23.307 $\pm$  0.129 &  24.395 $\pm$  0.140 & -15.716 $\pm$   0.065 \\
 SFF10-NB3-B12772 & SFACT J014259.69+274052.3 &    1:42:59.69 &   27:40:52.3 &  -0.52 $\pm$   0.09 &   5.57 & ELG &  22.981 $\pm$  0.126 &  22.067 $\pm$  0.063 &  24.609 $\pm$  0.216 & -15.558 $\pm$   0.063 \\
 SFF10-NB3-D13083 & SFACT J014300.87+280623.8 &    1:43:00.87 &   28:06:23.8 &  -1.18 $\pm$   0.10 &  11.59 & ELG &  22.741 $\pm$  0.127 &  22.534 $\pm$  0.099 &  23.547 $\pm$  0.107 & -15.414 $\pm$   0.032 \\
 SFF10-NB1-B12579 & SFACT J014300.97+274122.1 &    1:43:00.97 &   27:41:22.1 &  -0.52 $\pm$   0.08 &   6.62 & ELG &  22.108 $\pm$  0.070 &  21.398 $\pm$  0.040 &  22.774 $\pm$  0.058 & -15.518 $\pm$   0.050 \\
 SFF10-NB3-B12471 & SFACT J014301.74+273928.5 &    1:43:01.74 &   27:39:28.5 &  -0.61 $\pm$   0.11 &   5.39 & ELG &  23.218 $\pm$  0.123 &  22.704 $\pm$  0.078 &  23.677 $\pm$  0.085 & -15.722 $\pm$   0.064 \\
 SFF10-NB1-D12909 & SFACT J014302.15+281152.9 &    1:43:02.15 &   28:11:52.9 &  -0.45 $\pm$   0.04 &  10.02 & ExtG &  20.319 $\pm$  0.028 &  20.137 $\pm$  0.025 &  21.015 $\pm$  0.025 & -14.911 $\pm$   0.026 \\
 SFF10-NB3-B12244 & SFACT J014303.57+273655.2 &    1:43:03.57 &   27:36:55.2 &  -1.09 $\pm$   0.13 &   8.16 & ELG &  23.287 $\pm$  0.161 &  23.030 $\pm$  0.124 &  24.274 $\pm$  0.151 & -15.572 $\pm$   0.046 \\
 SFF10-NB2-B12225 & SFACT J014303.75+273744.4 &    1:43:03.75 &   27:37:44.4 &  -1.04 $\pm$   0.12 &   8.91 & ELG &  22.974 $\pm$  0.150 &  22.436 $\pm$  0.089 &  23.421 $\pm$  0.094 & -15.520 $\pm$   0.043 \\
\\
 SFF10-NB3-B12131 & SFACT J014304.20+275025.1 &    1:43:04.20 &   27:50:25.1 &  -0.87 $\pm$   0.04 &  22.34 & ELG &  20.993 $\pm$  0.030 &  20.699 $\pm$  0.025 &  21.551 $\pm$  0.026 & -14.830 $\pm$   0.016 \\
 SFF10-NB3-B12081 & SFACT J014304.52+275030.9 &    1:43:04.52 &   27:50:30.9 &  -0.59 $\pm$   0.03 &  18.35 & ELG &  20.137 $\pm$  0.024 &  19.879 $\pm$  0.019 &  20.772 $\pm$  0.021 & -14.802 $\pm$   0.015 \\
 SFF10-NB1-B12096 & SFACT J014304.64+273418.2 &    1:43:04.64 &   27:34:18.2 &  -0.59 $\pm$   0.09 &   6.66 & ELG &  22.348 $\pm$  0.087 &  21.572 $\pm$  0.046 &  22.685 $\pm$  0.054 & -15.453 $\pm$   0.057 \\
 SFF10-NB3-D12508 & SFACT J014305.18+275632.9 &    1:43:05.18 &   27:56:32.9 &  -0.70 $\pm$   0.12 &   5.66 & ELG &  23.594 $\pm$  0.281 &  22.564 $\pm$  0.111 &  23.301 $\pm$  0.099 & -15.670 $\pm$   0.073 \\
 SFF10-NB3-B11870 & SFACT J014306.06+274431.6 &    1:43:06.06 &   27:44:31.6 &  -0.75 $\pm$   0.13 &   5.68 & ELG &  23.412 $\pm$  0.145 &  22.970 $\pm$  0.093 &  24.131 $\pm$  0.116 & -15.742 $\pm$   0.064 \\
 SFF10-NB3-B11533 & SFACT J014307.68+275244.6 &    1:43:07.68 &   27:52:44.6 &  -0.48 $\pm$   0.09 &   5.36 & ELG &  22.565 $\pm$  0.100 &  22.222 $\pm$  0.077 &  22.888 $\pm$  0.065 & -15.623 $\pm$   0.049 \\
 SFF10-NB3-D12024 & SFACT J014308.50+281103.9 &    1:43:08.50 &   28:11:03.9 &  -0.64 $\pm$   0.12 &   5.16 & ELG &  23.389 $\pm$  0.117 &  22.811 $\pm$  0.075 &  24.237 $\pm$  0.106 & -15.860 $\pm$   0.063 \\
 SFF10-NB2-D11997 & SFACT J014308.81+281226.9 &    1:43:08.81 &   28:12:26.9 &  -0.88 $\pm$   0.13 &   6.71 & ELG &  22.853 $\pm$  0.129 &  23.009 $\pm$  0.170 &  23.441 $\pm$  0.097 & -15.653 $\pm$   0.062 \\
 SFF10-NB3-D11938 & SFACT J014309.34+280305.7 &    1:43:09.34 &   28:03:05.7 &  -1.41 $\pm$   0.09 &  16.23 & ELG &  21.941 $\pm$  0.079 &  21.785 $\pm$  0.066 &  22.297 $\pm$  0.051 & -15.029 $\pm$   0.032 \\
 SFF10-NB3-D11925 & SFACT J014309.51+275434.0 &    1:43:09.51 &   27:54:34.0 &  -0.47 $\pm$   0.09 &   5.42 & ELG &  22.771 $\pm$  0.137 &  21.581 $\pm$  0.046 &  22.368 $\pm$  0.045 & -15.727 $\pm$   0.057 \\
\\
 SFF10-NB3-B11195 & SFACT J014309.97+273813.9 &    1:43:09.97 &   27:38:13.9 &  -0.87 $\pm$   0.13 &   6.67 & ELG &  23.137 $\pm$  0.123 &  22.993 $\pm$  0.103 &  23.567 $\pm$  0.081 & -15.608 $\pm$   0.055 \\
 SFF10-NB3-D11772 & SFACT J014310.63+280641.6 &    1:43:10.63 &   28:06:41.6 &  -1.91 $\pm$   0.22 &   8.82 & ELG &  24.347 $\pm$  0.311 &  23.978 $\pm$  0.219 &  25.532 $\pm$  0.377 & -15.512 $\pm$   0.042 \\
 SFF10-NB1-B10986 & SFACT J014311.40+273300.7 &    1:43:11.40 &   27:33:00.7 &  -1.21 $\pm$   0.12 &   9.93 & ELG &  23.077 $\pm$  0.135 &  22.663 $\pm$  0.083 &  23.290 $\pm$  0.069 & -15.392 $\pm$   0.039 \\
 SFF10-NB1-B10674 & SFACT J014313.32+274124.6 &    1:43:13.32 &   27:41:24.6 &  -0.75 $\pm$   0.04 &  16.69 & ELG &  21.134 $\pm$  0.038 &  20.823 $\pm$  0.030 &  21.422 $\pm$  0.026 & -15.008 $\pm$   0.022 \\
 SFF10-NB2-B10675 & SFACT J014313.39+273356.3 &    1:43:13.39 &   27:33:56.3 &  -0.47 $\pm$   0.06 &   7.38 & ELG &  21.985 $\pm$  0.054 &  21.429 $\pm$  0.035 &  22.725 $\pm$  0.046 & -15.493 $\pm$   0.041 \\
 SFF10-NB1-D11121 & SFACT J014315.23+280125.9 &    1:43:15.23 &   28:01:25.9 &  -1.00 $\pm$   0.09 &  10.82 & ELG &  22.779 $\pm$  0.089 &  22.039 $\pm$  0.049 &  23.259 $\pm$  0.061 & -15.322 $\pm$   0.037 \\
 SFF10-NB3-D11032 & SFACT J014315.47+281521.9 &    1:43:15.47 &   28:15:21.9 &  -1.27 $\pm$   0.06 &  19.98 & ELG &  21.873 $\pm$  0.070 &  21.629 $\pm$  0.055 &  22.815 $\pm$  0.055 & -14.941 $\pm$   0.020 \\
 SFF10-NB3-D10890 & SFACT J014316.09+275845.4 &    1:43:16.09 &   27:58:45.4 &  -0.66 $\pm$   0.03 &  20.21 & HII &  20.058 $\pm$  0.020 &  19.625 $\pm$  0.015 &  20.703 $\pm$  0.018 & -14.659 $\pm$   0.015 \\
 SFF10-NB3-D10873 & SFACT J014316.14+275843.3 &    1:43:16.14 &   27:58:43.3 &  -0.38 $\pm$   0.02 &  17.35 & ExtG &  19.055 $\pm$  0.019 &  18.647 $\pm$  0.014 &  19.811 $\pm$  0.016 & -14.418 $\pm$   0.014 \\
 SFF10-NB3-B10141 & SFACT J014316.41+274430.5 &    1:43:16.41 &   27:44:30.5 &  -1.28 $\pm$   0.20 &   6.43 & ELG &  24.082 $\pm$  0.250 &  23.518 $\pm$  0.144 &  24.358 $\pm$  0.137 & -15.788 $\pm$   0.062 \\
  \enddata

\tablecomments{Table \ref{tab:SFF10} is published in its entirety in the machine-readable format. A portion is shown here for guidance regarding its form and content.}
\end{deluxetable*}



\begin{deluxetable*}{ccccccccccc}
    \tabletypesize{\scriptsize}
\rotate
\caption{SFF15 Emission-Line Objects \label{tab:SFF15}}
    \tablehead{
    \\
    SFACT Object ID & SFACT Coordinate ID & $\alpha$(J2000) & $\delta$(J2000) & $\Delta$m & ratio & Object & m$_r$ & m$_i$ & m$_g$ & log(f$_{NB}$)\\
    & & H:M:S & D:M:S & mag & & Type & mag & mag & mag &  erg/s/cm$^2$ \\  
    (1) & (2) & (3) & (4) & (5) & (6) & (7) & (8) & (9) & (10) & (11)
    }
 \startdata
 SFF15-NB3-B14284 & SFACT J023730.56+274425.1 &    2:37:30.56 &   27:44:25.1 &  -0.71 $\pm$   0.10 &   7.11 & ExtG &  21.747 $\pm$  0.068 &  21.774 $\pm$  0.075 &  22.427 $\pm$  0.062 & -15.413 $\pm$   0.048 \\
 SFF15-NB1-B14251 & SFACT J023730.82+274059.9 &    2:37:30.82 &   27:40:59.9 &  -0.94 $\pm$   0.17 &   5.63 & ELG &  23.294 $\pm$  0.178 &  22.758 $\pm$  0.113 &  23.926 $\pm$  0.134 & -15.741 $\pm$   0.065 \\
 SFF15-NB2-D24348 & SFACT J023731.19+281026.8 &    2:37:31.19 &   28:10:26.8 &  -0.68 $\pm$   0.13 &   5.14 & ELG &  22.734 $\pm$  0.119 &  22.840 $\pm$  0.147 &  23.688 $\pm$  0.127 & -15.849 $\pm$   0.094 \\
 SFF15-NB2-B14131 & SFACT J023731.68+272845.4 &    2:37:31.68 &   27:28:45.4 &  -1.29 $\pm$   0.21 &   5.99 & ELG &  23.378 $\pm$  0.223 &  23.925 $\pm$  0.355 &  24.420 $\pm$  0.252 & -15.731 $\pm$   0.089 \\
 SFF15-NB3-B14046 & SFACT J023731.96+275052.4 &    2:37:31.96 &   27:50:52.4 &  -0.47 $\pm$   0.07 &   6.65 & ELG &  22.050 $\pm$  0.047 &  21.716 $\pm$  0.041 &  22.908 $\pm$  0.053 & -15.560 $\pm$   0.042 \\
 SFF15-NB2-D22777 & SFACT J023733.00+281011.2 &    2:37:33.00 &   28:10:11.2 &  -1.53 $\pm$   0.04 &  34.51 & ELG &  21.096 $\pm$  0.040 &  21.435 $\pm$  0.057 &  21.788 $\pm$  0.037 & -14.644 $\pm$   0.016 \\
 SFF15-NB2-B13724 & SFACT J023734.14+274128.0 &    2:37:34.14 &   27:41:28.0 &  -0.79 $\pm$   0.07 &  11.71 & ELG &  22.404 $\pm$  0.047 &  22.260 $\pm$  0.045 &  23.404 $\pm$  0.058 & -15.522 $\pm$   0.030 \\
 SFF15-NB2-B13721 & SFACT J023734.15+274129.1 &    2:37:34.15 &   27:41:29.1 &  -0.90 $\pm$   0.06 &  16.11 & ELG &  21.621 $\pm$  0.034 &  21.529 $\pm$  0.033 &  22.618 $\pm$  0.041 & -15.154 $\pm$   0.021 \\
 SFF15-NB2-D22292 & SFACT J023734.35+281003.1 &    2:37:34.35 &   28:10:03.1 &  -0.49 $\pm$   0.08 &   5.74 & ELG &  21.645 $\pm$  0.074 &  21.500 $\pm$  0.074 &  22.790 $\pm$  0.087 & -15.410 $\pm$   0.051 \\
 SFF15-NB2-D22224 & SFACT J023734.52+280514.9 &    2:37:34.52 &   28:05:14.9 &  -1.37 $\pm$   0.14 &   9.85 & ELG &  23.120 $\pm$  0.138 &  23.295 $\pm$  0.166 &  24.012 $\pm$  0.138 & -15.597 $\pm$   0.037 \\
\\
 SFF15-NB3-D22277 & SFACT J023734.53+275653.7 &    2:37:34.53 &   27:56:53.7 &  -0.49 $\pm$   0.08 &   5.98 & ELG &  22.342 $\pm$  0.061 &  21.809 $\pm$  0.042 &  23.564 $\pm$  0.077 & -15.633 $\pm$   0.048 \\
 SFF15-NB3-B13580 & SFACT J023735.23+274121.9 &    2:37:35.23 &   27:41:21.9 &  -0.48 $\pm$   0.05 &   9.76 & ELG &  21.423 $\pm$  0.031 &  21.181 $\pm$  0.026 &  21.581 $\pm$  0.022 & -15.288 $\pm$   0.024 \\
 SFF15-NB2-D20719 & SFACT J023738.93+280530.2 &    2:37:38.93 &   28:05:30.2 &  -0.52 $\pm$   0.08 &   6.34 & ELG &  21.875 $\pm$  0.059 &  22.076 $\pm$  0.072 &  22.966 $\pm$  0.065 & -15.633 $\pm$   0.035 \\
 SFF15-NB1-B12855 & SFACT J023740.25+274942.1 &    2:37:40.25 &   27:49:42.1 &  -0.80 $\pm$   0.10 &   7.79 & ELG &  22.611 $\pm$  0.100 &  22.174 $\pm$  0.072 &  22.864 $\pm$  0.071 & -15.550 $\pm$   0.049 \\
 SFF15-NB2-B12874 & SFACT J023740.34+272939.9 &    2:37:40.34 &   27:29:39.9 &  -1.17 $\pm$   0.09 &  13.00 & ELG &  22.549 $\pm$  0.070 &  22.583 $\pm$  0.078 &  23.594 $\pm$  0.078 & -15.367 $\pm$   0.029 \\
 SFF15-NB2-B12729 & SFACT J023741.60+272959.4 &    2:37:41.60 &   27:29:59.4 &  -1.70 $\pm$   0.14 &  11.94 & ELG &  22.939 $\pm$  0.159 &  23.720 $\pm$  0.345 &  23.689 $\pm$  0.161 & -15.291 $\pm$   0.033 \\
 SFF15-NB2-D19839 & SFACT J023741.93+281035.0 &    2:37:41.93 &   28:10:35.0 &  -0.63 $\pm$   0.12 &   5.11 & ELG &  22.770 $\pm$  0.097 &  22.906 $\pm$  0.142 &  23.588 $\pm$  0.105 & -15.757 $\pm$   0.066 \\
 SFF15-NB1-B12675 & SFACT J023742.16+272938.9 &    2:37:42.16 &   27:29:38.9 &  -0.74 $\pm$   0.13 &   5.60 & ELG &  22.927 $\pm$  0.100 &  22.736 $\pm$  0.083 &  24.033 $\pm$  0.114 & -15.662 $\pm$   0.071 \\
 SFF15-NB3-B12608 & SFACT J023742.63+274616.0 &    2:37:42.63 &   27:46:16.0 &  -0.51 $\pm$   0.09 &   5.56 & ELG &  22.604 $\pm$  0.058 &  22.288 $\pm$  0.049 &  23.423 $\pm$  0.065 & -15.707 $\pm$   0.054 \\
 SFF15-NB3-B12427 & SFACT J023744.04+274533.7 &    2:37:44.04 &   27:45:33.7 &  -0.63 $\pm$   0.05 &  12.28 & ELG &  21.254 $\pm$  0.034 &  21.076 $\pm$  0.029 &  22.018 $\pm$  0.032 & -15.248 $\pm$   0.024 \\
 \\
 SFF15-NB3-D18788 & SFACT J023744.26+281330.5 &    2:37:44.26 &   28:13:30.5 &  -0.54 $\pm$   0.08 &   7.05 & ELG &  22.009 $\pm$  0.051 &  21.968 $\pm$  0.055 &  22.820 $\pm$  0.054 & -15.391 $\pm$   0.046 \\
 SFF15-NB2-B12371 & SFACT J023744.60+273441.2 &    2:37:44.60 &   27:34:41.2 &  -0.60 $\pm$   0.06 &   9.79 & ExtG &  21.934 $\pm$  0.045 &  21.494 $\pm$  0.035 &  22.785 $\pm$  0.045 & -15.335 $\pm$   0.030 \\
 SFF15-NB1-D18539 & SFACT J023744.78+281340.3 &    2:37:44.78 &   28:13:40.3 &  -1.37 $\pm$   0.10 &  13.44 & ELG &  21.511 $\pm$  0.033 &  21.295 $\pm$  0.030 &  23.294 $\pm$  0.083 & -15.386 $\pm$   0.033 \\
 SFF15-NB3-B12212 & SFACT J023745.92+273918.5 &    2:37:45.92 &   27:39:18.5 &  -0.77 $\pm$   0.09 &   8.86 & ELG &  22.657 $\pm$  0.075 &  22.226 $\pm$  0.056 &  23.587 $\pm$  0.078 & -15.525 $\pm$   0.030 \\
 SFF15-NB3-D17352 & SFACT J023746.87+280421.5 &    2:37:46.87 &   28:04:21.5 &  -0.71 $\pm$   0.11 &   6.15 & ELG &  24.036 $\pm$  0.220 &  22.660 $\pm$  0.069 &  26.318 $\pm$  0.717 & -15.750 $\pm$   0.061 \\
 SFF15-NB3-B11965 & SFACT J023747.82+274851.5 &    2:37:47.82 &   27:48:51.5 &  -0.75 $\pm$   0.10 &   7.21 & ELG &  22.912 $\pm$  0.115 &  22.340 $\pm$  0.069 &  23.381 $\pm$  0.075 & -15.570 $\pm$   0.057 \\
 SFF15-NB1-D15218 & SFACT J023748.24+281511.6 &    2:37:48.24 &   28:15:11.6 &  -0.53 $\pm$   0.10 &   5.22 & ELG &  21.954 $\pm$  0.109 &  21.542 $\pm$  0.083 &  23.071 $\pm$  0.165 & -15.326 $\pm$   0.067 \\
 SFF15-NB1-D14190 & SFACT J023749.21+280902.6 &    2:37:49.21 &   28:09:02.6 &  -0.73 $\pm$   0.14 &   5.01 & ELG &  22.798 $\pm$  0.126 &  22.347 $\pm$  0.109 &  23.150 $\pm$  0.100 & -15.754 $\pm$   0.081 \\
 SFF15-NB2-B11684 & SFACT J023750.76+272858.2 &    2:37:50.76 &   27:28:58.2 &  -1.91 $\pm$   0.17 &  11.41 & ELG &  23.082 $\pm$  0.127 &  24.305 $\pm$  0.440 &  25.051 $\pm$  0.315 & -15.383 $\pm$   0.035 \\
 SFF15-NB3-D13288 & SFACT J023751.99+275346.3 &    2:37:51.99 &   27:53:46.3 &  -0.65 $\pm$   0.07 &   8.78 & ELG &  22.373 $\pm$  0.070 &  21.789 $\pm$  0.044 &  23.308 $\pm$  0.072 & -15.364 $\pm$   0.042 \\
   \enddata

\tablecomments{Table \ref{tab:SFF15} is published in its entirety in the machine-readable format. A portion is shown here for guidance regarding its form and content.}
\end{deluxetable*}

\section{Results}\label{Sec:results}
\subsection{SFACT Survey Catalogs}

We present the full list of our SFACT sources from our pilot-study fields in Tables~\ref{tab:SFF01} --~\ref{tab:SFF15}. In each table, column (1) is the SFACT ID, the unique identifier by which we refer to any candidate. This ID is made up of the field name (ex:\ SFF01), the filter designation (ex:\ NB3), the quadrant in which the object was found (ex:\ D), and a running number which is assigned in the initial object detection stage (ex:\ 20110). Together these quantities form the SFACT ID  SFF01-NB3-D20110.  We use the SFACT object ID to refer to specific sources throughout the remainder of this paper Column (2) provides an alternate coordinate-based designation, using IAU-approved nomenclature. Columns (3) and (4) give the astrometric positions of the object in J2000 coordinates on the Gaia astrometric system.  Comparison of the coordinates of stars found in the SFACT images with those cataloged in the SDSS shows that there is little or no systematic offsets between the two sets of coordinates (mean $\Delta\alpha$ and $\Delta\delta$ $\leq$ 0.05 arcsec for each field), and that the RMS scatter for individual stars is $\sim$0.15--0.20 arcsec.

Columns (5) and (6) give the $\Delta m$ and {\it ratio} values used to select candidates (see Section~\ref{ssec:indentifying}). 
Column (7) is the type of object, which is assigned during the selection process.  Three types of objects are identified: \HII regions in an extended galaxy (marked as HII), the centers of any extended galaxy (ExtG),  and generic emission-line objects (ELG).   A detailed explanation of these types is given below.  
Columns (8) through (10) are the broad-band magnitudes measured in the r, i, and g filters, respectively, along with their formal uncertainties.  
Finally, the emission-line flux in the relevant narrow-band filter is tabulated in column (11). 

The tables are sorted by the RA order of the objects within each field. All magnitudes and fluxes are the observed values.  No corrections for Galactic absorption have been applied.   Additionally, no corrections have been applied to the emission-line flux listed in column (11), either for any additional emission lines that might be present in the NB filter (e.g., [\ion{N}{2}] emission) or for bandpass corrections.  These corrections will be applied whenever appropriate in any subsequent analysis of the survey data.

SFACT is designed to be a comprehensive survey for extragalactic objects with emission lines.  We detect many nearby, resolved galaxies via their individual \HII regions, more distant galaxies via their ``global" emission, as well as unresolved sources of emission.  At the stage of creating the catalogs of our emission-line objects we do not possess any spectroscopic follow-up information, so all compact or unresolved emission-line candidates are labeled simply as ELG (even when subsequent spectroscopy reveals them to be QSOs).  As is detailed in SFACT1 and mentioned in Section~\ref{ssec:indentifying}, for objects detected via their disk \HII regions we will only catalog the brightest emission region present.  These objects are labeled as HII objects in column 7 of Tables~\ref{tab:SFF01} --~\ref{tab:SFF15}.   However, for the purpose of measuring the total emission-line flux (for total star-formation rates) and the total systemic BB photometry, we always catalog the central position of all galaxies that were detected via their disk \HII regions.  These objects are labeled as ExtG objects (for extended galaxies) in column 7.  

We stress that the ExtG objects {\it are} ELGs, but for the purposes of our survey methodology we need to distinguish them from the more generic ELGs and from the \HII regions.  They do not represent a new or different class of object.  All HII objects in our catalogs will have a corresponding ExtG catalog entry.   However, not all ExtG objects will have a related HII object if the line emission at the center of the galaxy is the strongest emitting region in the galaxy.   There are 474 objects labeled as ELG, 40 labeled as ExtG, and 19 labeled as HII in Tables~\ref{tab:SFF01} --~\ref{tab:SFF15}.

There are 533 total SFACT objects in these three pilot-study fields. A total of 1.50 deg$^2$ on the sky was searched.  Counting only unique objects (i.e., not double counting the 19 \HII regions and the corresponding galaxy centers), this gives us a surface density of 342.7 SFACT objects deg$^{-2}$. 

\subsection{Example Objects}
We illustrate the types of objects detected in our survey by showing ten examples of SFACT candidates. These are all objects which were selected for follow-up spectroscopy and have been confirmed to be real detections. We have chosen examples which demonstrate the variety of objects found in the SFACT catalog and the depth of our images. The example objects have been grouped by their detected line. For each object, the redshift and type of object is derived from spectral analysis. This is discussed in SFACT3 where the corresponding spectra for these example objects can be found.  These image cutouts are produced by our software and are used during ELG candidate evaluations and checking.

\begin{figure}[htp]
\subfloat{%
  \includegraphics[clip,width=\columnwidth]{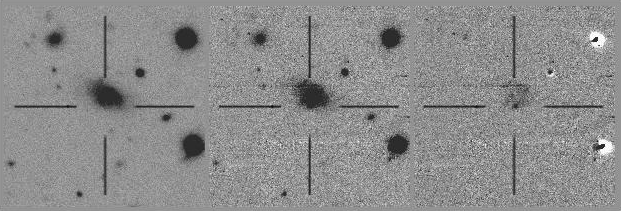}%
}

\subfloat{%
  \includegraphics[clip,width=\columnwidth]{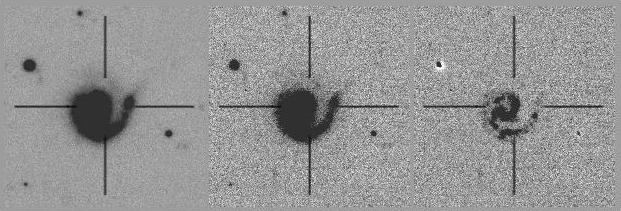}%
}

\subfloat{%
  \includegraphics[clip,width=\columnwidth]{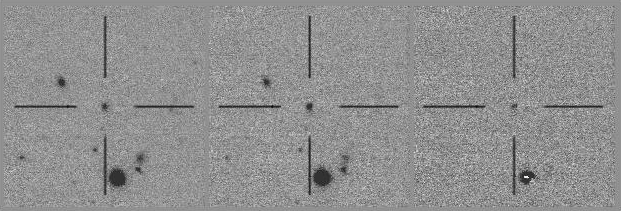}%
}
\caption{Three H$\alpha$-detected SFACT objects.  Each row shows three 50\arcsec $\times$ 50\arcsec\ image cutouts of the continuum image (left), the NB image (middle), and the difference image (right).  The \HA redshift windows detect objects in the range 0.0 $<$ z $<$ 0.15 and include all of our extended galaxies. Top: SFF01-NB2-B19198 was detected in the NB2 filter and is a low-luminosity dwarf galaxy with z = 0.0034. Middle: SFF15-NB1-A2606 was detected in the NB1 filter via its many \HII regions (z = 0.0643). Bottom: SFF01-NB3-D2175 was detected in the NB3 filter and has z = 0.1374.}
\label{Fig:HA detections}
\end{figure}

\begin{figure}[htp]
\subfloat{%
  \includegraphics[clip,width=\columnwidth]{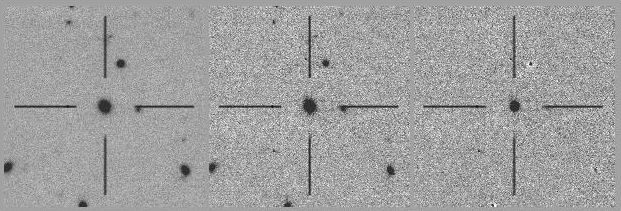}%
}

\subfloat{%
  \includegraphics[clip,width=\columnwidth]{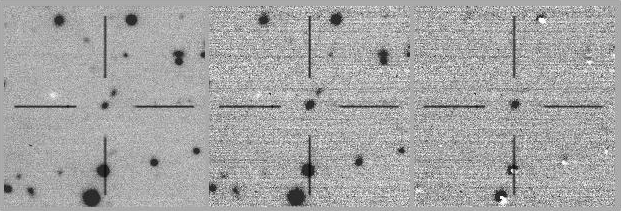}%
}

\subfloat{%
  \includegraphics[clip,width=\columnwidth]{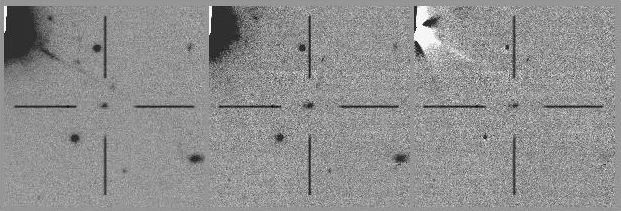}%
}
\caption{Three [\ion{O}{3}]-detected objects.  Each row shows three 50\arcsec $\times$ 50\arcsec\ image cutouts of the continuum image (left), the NB image (middle), and the difference image (right).  The  [\ion{O}{3}] redshift windows detect objects in the range 0.31 $<$ z $<$ 0.50, and the [\ion{O}{3}]-detected objects are  typically compact sources like these. Top: SFF15-NB2-C20849 was detected in the NB2 filter; it is a Seyfert 2 galaxy at z = 0.3228. Middle: SFF01-NB1-D4500 was detected in the NB1 filter and is a star-forming galaxy at z = 0.3906. Bottom: SFF10-NB3-D13569 was detected in the NB3 filter at z = 0.4829.}
\label{Fig:OIII detections}
\end{figure}

\begin{figure}[htp]
\subfloat{%
  \includegraphics[clip,width=\columnwidth]{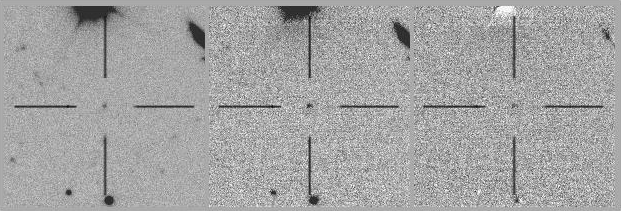}%
}

\subfloat{%
  \includegraphics[clip,width=\columnwidth]{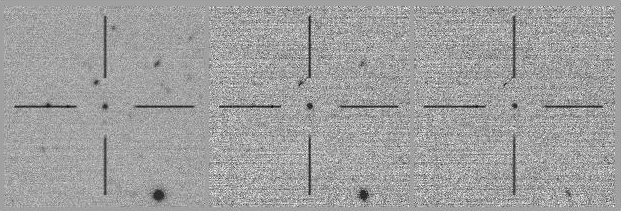}%
}

\subfloat{%
  \includegraphics[clip,width=\columnwidth]{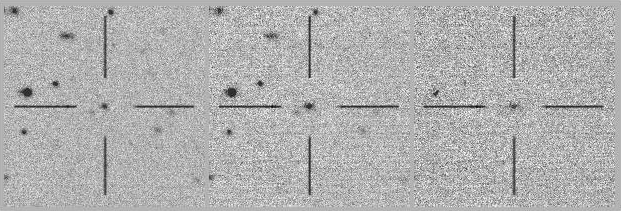}%
}
\caption{Three [\ion{O}{2}]-detected objects.  Each row shows three 50\arcsec $\times$ 50\arcsec\ image cutouts of the continuum image (left), the NB image (middle), and the difference image (right).  The  [\ion{O}{2}] redshift windows detect objects in the range 0.78 $<$ z $<$ 1.0 and are typically small dots like these. Top: SFF10-NB2-A8098 was detected in the NB2 filter (z = 0.7670). Middle: SFF10-NB1-C19716 was detected in the NB1 filter and has z = 0.8694.  Bottom: SFF01-NB3-B5847 was found in the NB3 filter (z = 1.0023).}
\label{Fig:OII detections}
\end{figure}

\begin{figure}
    \centering
    \includegraphics[width=.8\paperwidth]{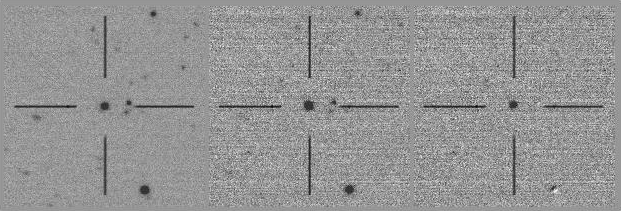}
    \caption{SFF10-NB2-C21205.  Shown are three 50\arcsec $\times$ 50\arcsec\ image cutouts of the continuum image (left), the NB image (middle), and the difference image (right).  This object was detected due its strong \ion{C}{3}] line at 1908\AA\ falling in the NB2 filter. Follow-up spectra reveal it to be a quasar at z = 2.464.}
    \label{find:C21205}
\end{figure}

\subsubsection{\HA Detections}
The first three example SFACT galaxies, shown in Figure~\ref{Fig:HA detections}, were all detected via their \HA emission line.  SFF01-NB2-B19198 at the top of Figure~\ref{Fig:HA detections} is one of our closest ELGs at $z=0.0034$. The specific object is not actually the galaxy center, but an \HII region near the center. As discussed in Section~\ref{ssec:apphot}, the \HII region remains in our catalog, but the photometric properties measured are those for the galaxy as a whole. Here it is visually clear that the \HII region is a large knot of emission in an otherwise quiescent dwarf galaxy. In a more traditional BB-only survey this may not have stood out as a source of emission. This galaxy has a total g-band magnitude of 19.00 and a narrow-band flux of 2.24 \fluxunitV. 

The middle galaxy is SFF15-NB1-A2606, which was detected in our NB1 filter.  Again, this is an \HII region in a larger galaxy. 
This spiral galaxy is found at $z=0.0643$ with a g-band magnitude of 17.16 and an integrated narrow-band flux of 2.36 \fluxunitIV. Both of these first two galaxies demonstrate the ability of SFACT to find \HII regions in extended sources. 

The last of this set is a more typical SFACT ELG. SFF01-NB3-D2175 is a compact object which is visible in the continuum image and appears slightly brighter in the NB image. This particular galaxy has a g-band magnitude of 22.41, a NB flux of 2.80 \fluxunitVI, and is found at $z=0.1374$.  This system is a low-luminosity dwarf star-forming galaxy.

\subsubsection{\OIII Detections}

The next three examples, shown in Figure~\ref{Fig:OIII detections}, are each \OIII detections. In the top set of images is SFF15-NB2-C20849. This galaxy has very strong line emission. In our nearest \OIII redshift window, this object is at $z=0.3228$ with a g-band magnitude of 21.17 and narrow-band flux of 6.81 \fluxunitV, making it the object with the second strongest flux in this example set, and the strongest of those which are not large spiral galaxies. Follow-up analysis (discussed in SFACT3) has confirmed that this object is a Seyfert 2.

The middle set of images shows SFF01-NB1-D4500 which has a g-band magnitude of 22.56. This object is found at $z=0.3906$ and has a NB flux of 1.01 \fluxunitV.  This object represents a strong, clear detection and may be a Green Pea-like star-forming galaxy \citep[e.g.,][]{gp, brunker2020}.

Rounding out the [\ion{O}{3}]-detected set is SFF10-NB3-D13569 at the bottom of Figure~\ref{Fig:OIII detections}. This system is at a redshift of $z=0.4829$ with a g-band magnitude of 23.02 and a narrow-band flux of 5.03 \fluxunitVI.  This object is representative of many SFACT objects for which the NB flux is not overwhelmingly strong yet still have a strong enough emission line for us to clearly identify it as an object of interest.

\begin{figure}
    \makebox[\textwidth]{\includegraphics[width=.84\paperwidth]{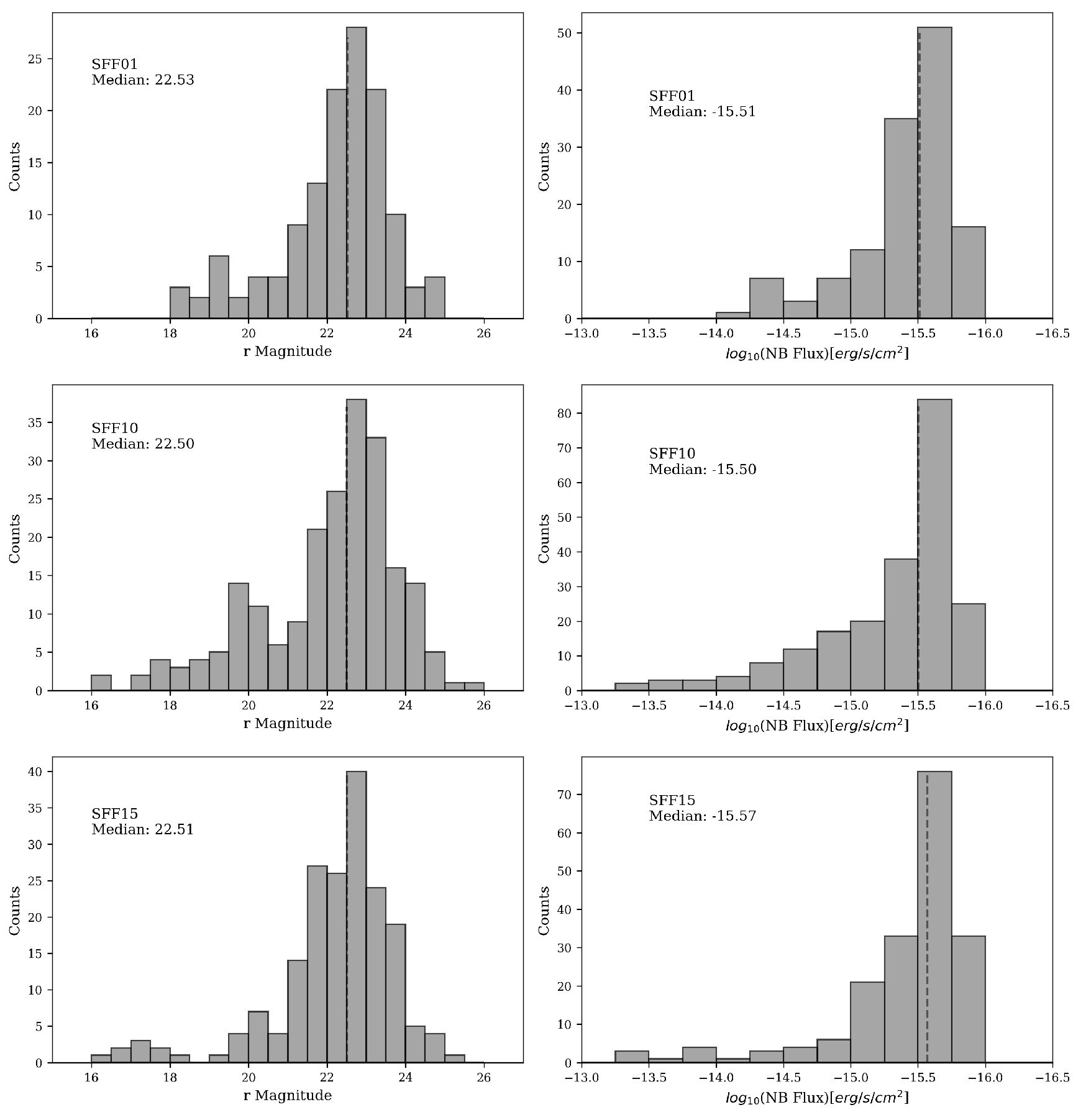}}
    \caption{Distributions of r magnitude and NB flux broken down by field. The left hand plots show the r magnitude distributions for each of the pilot-study fields while the right shows the NB flux distributions. From top to bottom is SFF01, SFF10, then SFF15. The vertical dashed lines mark the median of each distribution. The distributions are seen to be very similar from field to field.}\label{fig:CompositeHists}
\end{figure}

\begin{figure}
    \makebox[\textwidth]{\includegraphics[width=.84\paperwidth]{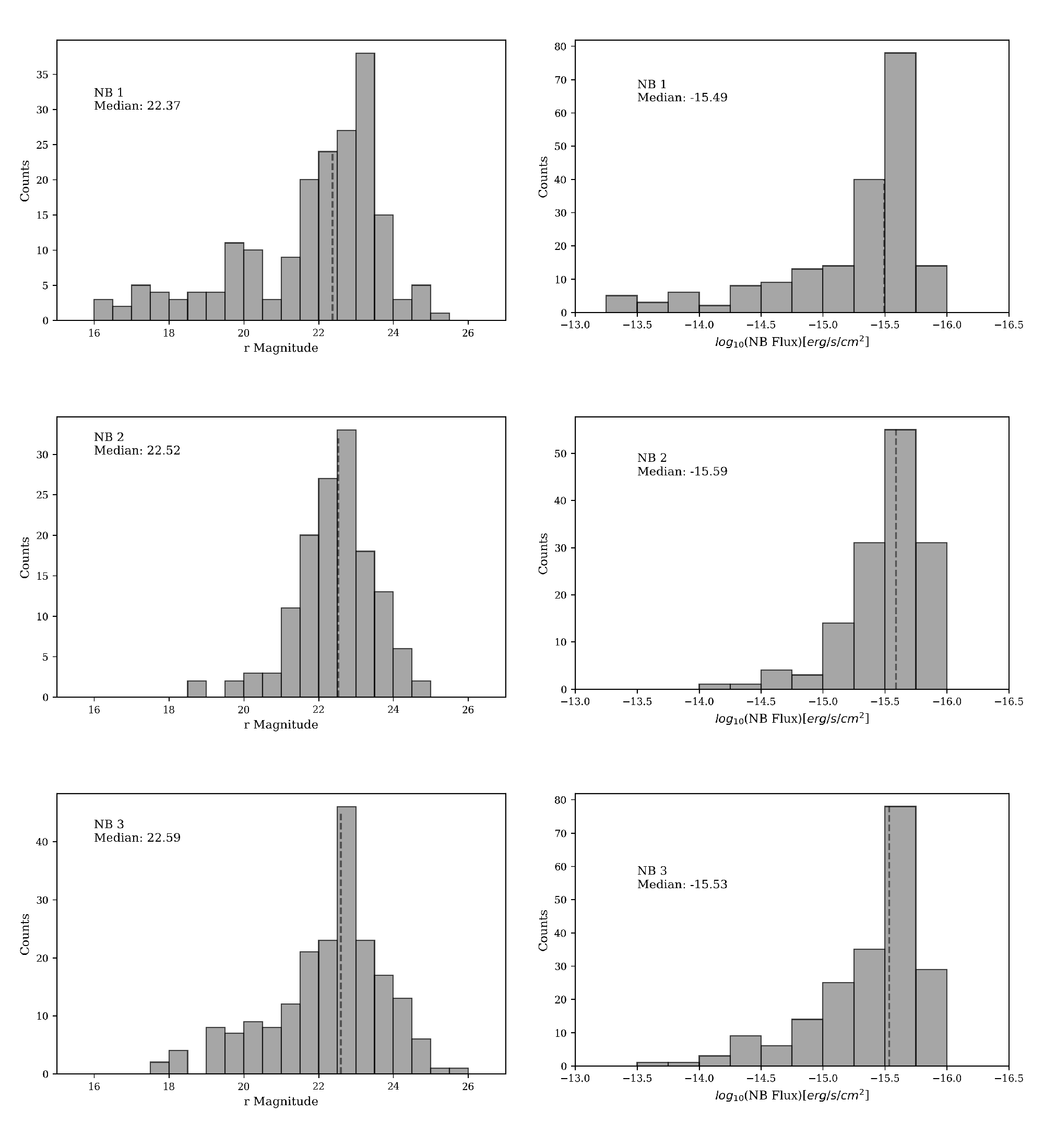}}
    \caption{Distributions of r magnitude and NB flux broken down by NB filter. The left hand plots show the r magnitude distributions for each of the pilot-study fields while the right shows the NB flux distributions. From top to bottom is NB1, NB2, then NB3. The vertical dashed lines mark the median of each distribution. Again, there is a strong similarity between the distributions from the different filters, but with a few notable differences.}\label{fig:FilterHists}
\end{figure}

\begin{figure}
    \centering
    \includegraphics[scale=1]{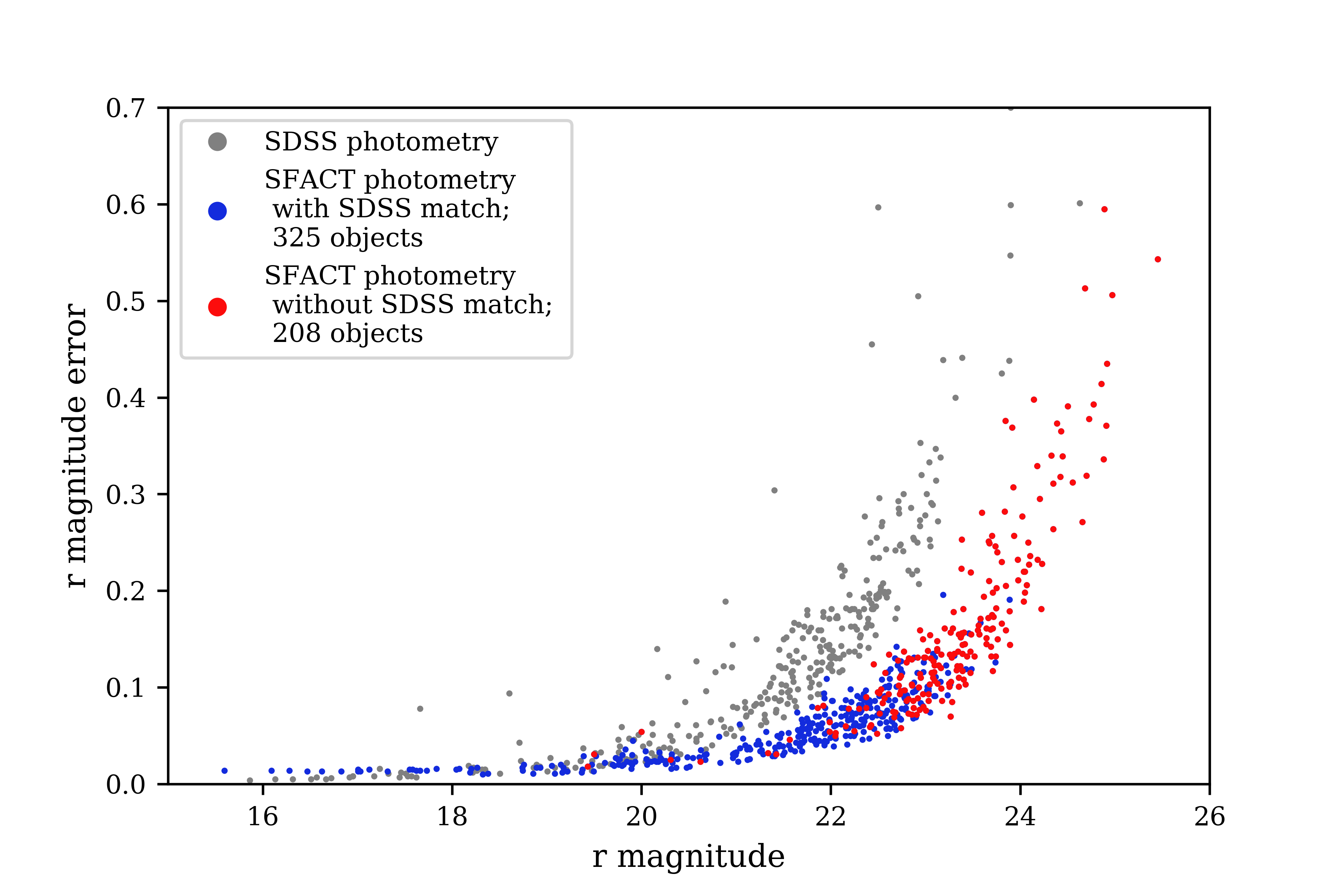}
    \caption{A plot of the photometric uncertainty vs. r-band magnitude for all 533 SFACT pilot-study objects.  SFACT objects that are included in the SDSS catalog are shown in blue, while those that are not detected by SDSS are plotted in red.   SDSS photometry for objects in common is shown in gray.  The SFACT photometry is seen to have good fidelity to r $\sim$ 24.}
    \label{fig:rmagcompare}
\end{figure}

\subsubsection{\OII Detections}

The final set of example objects are shown in Figure~\ref{Fig:OII detections} and were each detected by their \OII line. SFF10-NB2-A8098 is shown in the top set of Figure~\ref{Fig:OII detections} and is one of our fainter sources at a g-band magnitude of 23.78, falling over half a magnitude below the median g-band magnitude of the pilot-study objects (SFACT1). This demonstrates the sensitivity of SFACT. The galaxy in the NB image before continuum subtraction (the middle cutout) looks brighter than in the continuum image on the left, demonstrating the visually strong emission line. It is at a distance of $z=0.7670$ and has a narrow-band flux of 2.04 \fluxunitVI.  

On the middle row is SFF10-NB1-C19716. This galaxy is at $z=0.8694$; at such a distance it is understandably very compact in our images. This object has a g-band magnitude of 23.06 and a narrow-band flux of 3.63 \fluxunitVI.

One of the most distant galaxies in our primary redshift windows is SFF01-NB3-B5847 at $z=1.0023$. It has a g-band magnitude of 23.08 and a narrow-band flux of 4.51 \fluxunitVI. 

\subsubsection{Other Detections}

SFACT detects objects outside of our primary redshift windows, including numerous QSOs. The last example object, shown in Figure~\ref{find:C21205}, is one such QSO.   For this object, the \ion{C}{3}] emission line at 1908~\AA\ is redshifted into our NB2 filter, allowing us to detect it. As can be seen from Figure~\ref{find:C21205}, it is a bright target, with a g-band magnitude of 20.95. It exhibits a moderate line flux of 7.94 \fluxunitVI\ yet its distance, $z=2.4643$, demonstrates SFACT's ability to detect objects well beyond $z=1$. There are a total of 13 objects in this pilot study which are at redshifts greater than our primary redshift windows, all of which have been verified as QSOs in our follow-up spectroscopic observations (see SFACT3).

The spectra corresponding to all of the example SFACT objects shown here can be found in SFACT3.

\vfill\eject
\subsection{Photometric Properties of SFACT Objects}
In this section we examine the photometric properties of the SFACT objects.  SFACT1 (Figure 2) presents a set of composite histograms showing the range of BB apparent magnitudes for the full sample of pilot-study candidates, demonstrating the depth of our sample. In this paper, we examine these photometric properties in more detail by comparing the distributions of BB magnitude and NB line flux across the three pilot-study fields as well as across the three narrow-band filters. 

Figure~\ref{fig:CompositeHists} shows the distribution of r-band magnitude and NB flux for each pilot-study field separately. While there are variations between the fields, the broad characteristics are very similar. The median r-band magnitudes are 22.53, 22.50 and 22.51, demonstrating a remarkably consistent depth between the fields. This figure also demonstrates the range of brightnesses in our catalog. We see objects which have an r-band magnitude as bright as 16 and as faint as 25. The NB flux distribution on the right hand side of Figure~\ref{fig:CompositeHists} also exhibits strong similarities across the survey fields. The median log NB flux is seen to be very stable across the three fields: -15.51, -15.50, and -15.57 \logfluxunit. 

Figure 2 of SFACT1 also presents a composite g-r histogram. Like the BB magnitudes, there is a broad range of colors represented in the sample. The median g-r color of 0.65 is consistent with early-type spiral galaxies, but the bulk of the sample have colors between $0.2<g-r<1.2$ and include many red systems. As discussed in SFACT1, this is due in part to our selection method. Strong emission lines are present in many of our candidates, and these strong lines can influence the overall color of the galaxy, leading to an actively star-forming system appearing redder than expected \citep[e.g.,][]{gp, yang2017}. These strong emission lines can be seen as part of a wide range of emission line strengths in Figure~\ref{fig:CompositeHists}. This figure highlights SFACT's sensitivity. The strong peak in log(f$_{NB}$) between -15.50 and -15.75 implies that our survey is complete to approximately this level. 

As another way of viewing the distribution, Figure~\ref{fig:FilterHists} shows a similar set of histograms, this time broken down according to which NB filter the object was detected in. While there is a slightly greater spread in the median values, there is still strong consistency across the data sets. The most striking difference is the extended bright end of the distributions in NB1 and the deficit of brighter sources in NB2. The latter is presumably caused by the almost complete lack of \HA detections in NB2. This is expected, due to the limited volume over which any \HA sources could be found within the NB2 filter. Conversely, NB1 finds more H$\alpha$-detected galaxies that are bright.

Figure~\ref{fig:rmagcompare} presents a plot of photometric error vs. r-band magnitude for all 533 SFACT candidates.  Out of the complete sample, 325 (61.0\%) are detected in the SDSS database; these objects are plotted as blue dots, while the corresponding SDSS r-band photometry is plotted as gray dots.  The remaining 208 SFACT objects (39.0\%), plotted as red dots, are too faint to have been detected in SDSS.   The differences in the error curves between SFACT and SDSS are expected, since SFACT is carried out on a larger telescope and employs longer effective exposure times than SDSS.

The key point presented in Figure~\ref{fig:rmagcompare} is the high quality of the error curve for the SFACT photometry.   The median value of $\sigma_r$ for objects with r = 22.0 $\pm$ 0.2 is 0.061 mag.  For SFACT sources with r $\sim$ 23.0 the median uncertainty is 0.104 mag, while at r $\sim$ 24.0 the median value of $\sigma_r$ is 0.229 mag.  The SFACT photometry yields high-integrity measurements well beyond the median r-band magnitude of the sample (i.e., r $\sim$ 22.5).

\subsection{Connecting Selection Parameters to NB Flux}

\begin{figure}[t]
    \centering
    \includegraphics[scale=1]{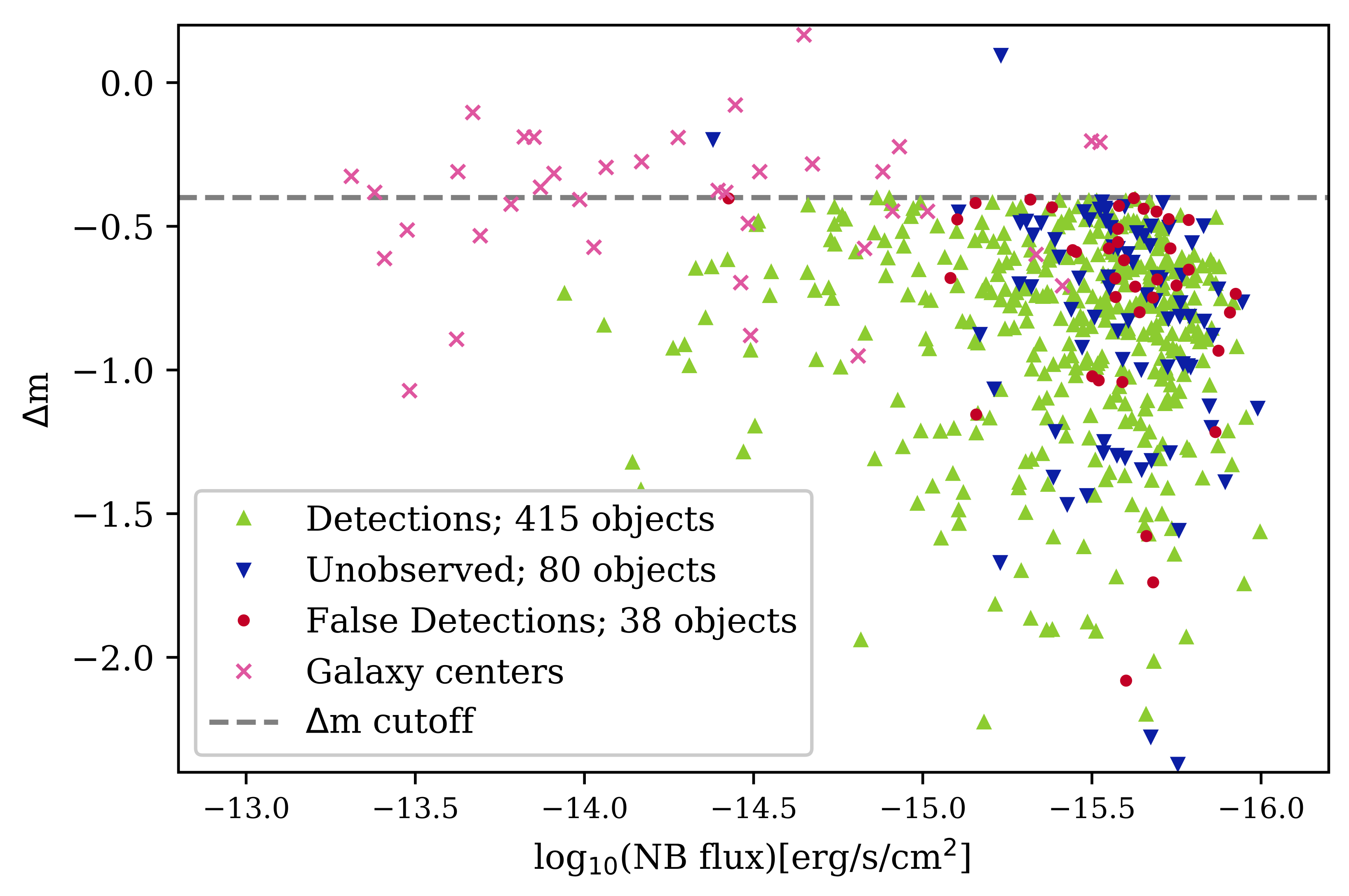}
    \caption{Shown here are the SFACT objects comparing their $\Delta m$ against their measured NB flux. The dashed vertical line shows the cutoff of objects which proceed to the next step of selection process. Objects marked as pink crosses are galaxy centers. Blue downward triangles are candidates which do not yet have follow-up spectroscopy. Green upward triangles have been confirmed as ELGs and red circles denote objects which are confirmed to be false detections. The sample size of each is indicated in the legend.}
    \label{fig:MagdiffVnbflux}
\end{figure}

In this section we investigate how well our measured emission-line fluxes correlate with the target selection parameters we presented earlier in Figure~\ref{fig:diag}. In  Figure~\ref{fig:MagdiffVnbflux} we show the $\Delta m$ values and the corresponding NB flux for each object. Because of our follow-up spectroscopy (discussed in SFACT3) we are able to denote confirmed emission-line objects (green upward triangles) and false detections (red circles) while also marking those which have yet to be observed (blue downward triangles). The dashed line marks the $\Delta m$ cutoff of -0.4 as one of our selection criteria to identify ELG candidates. Anything above this line is a galaxy center (ExtG in Tables~\ref{tab:SFF01} --~\ref{tab:SFF15}); these objects will always have an \HII region located somewhere below the cutoff line\footnote{We note that, because of the way they are selected, our ExtG objects will not necessarily satisfy both of our selection criteria.  The values of $\Delta m$ and {\it ratio} for these objects are those associated with the galaxy center, which in some cases does not emit significant line flux.  Hence, many of these sources will be located outside the ranges denoted by our selection limits in the plots shown in this section.  These objects are valid SFACT objects since one or more \HII regions found within the galaxy do satisfy our selection criteria.}. There is no strong correlation between $\Delta m$ and the strength of the emission line. This is expected since $\Delta m$ is a flux  ratio which should not scale with an absolute flux. 

Rather, we expect the strongest correlation to be between $\Delta m$ and the emission-line equivalent width (EW). Since $\Delta m$ is a measure of excess flux in the difference image, we expect that a larger excess is driven by stronger line emission.  However, as explained in SFACT3, our spectroscopic EW measurements are not all reliable. This is due to the sky-subtraction procedure followed for our multi-fiber spectra combined with the extremely faint nature of many of our objects. Our sky subtraction often over-subtracts the continuum by small amounts, leading to slightly negative continuum measurements for some of our faintest sources and resulting in indeterminate EWs. Even when the continuum is positive, this effect can result in unphysically large EWs (e.g., EW$_{5007} >$ 2000\AA). While the majority of our spectroscopic EWs appear to be reliably measured, the outliers render our EWs dubious and undependable.

\begin{figure}[t]
    \centering
    \includegraphics[scale=1]{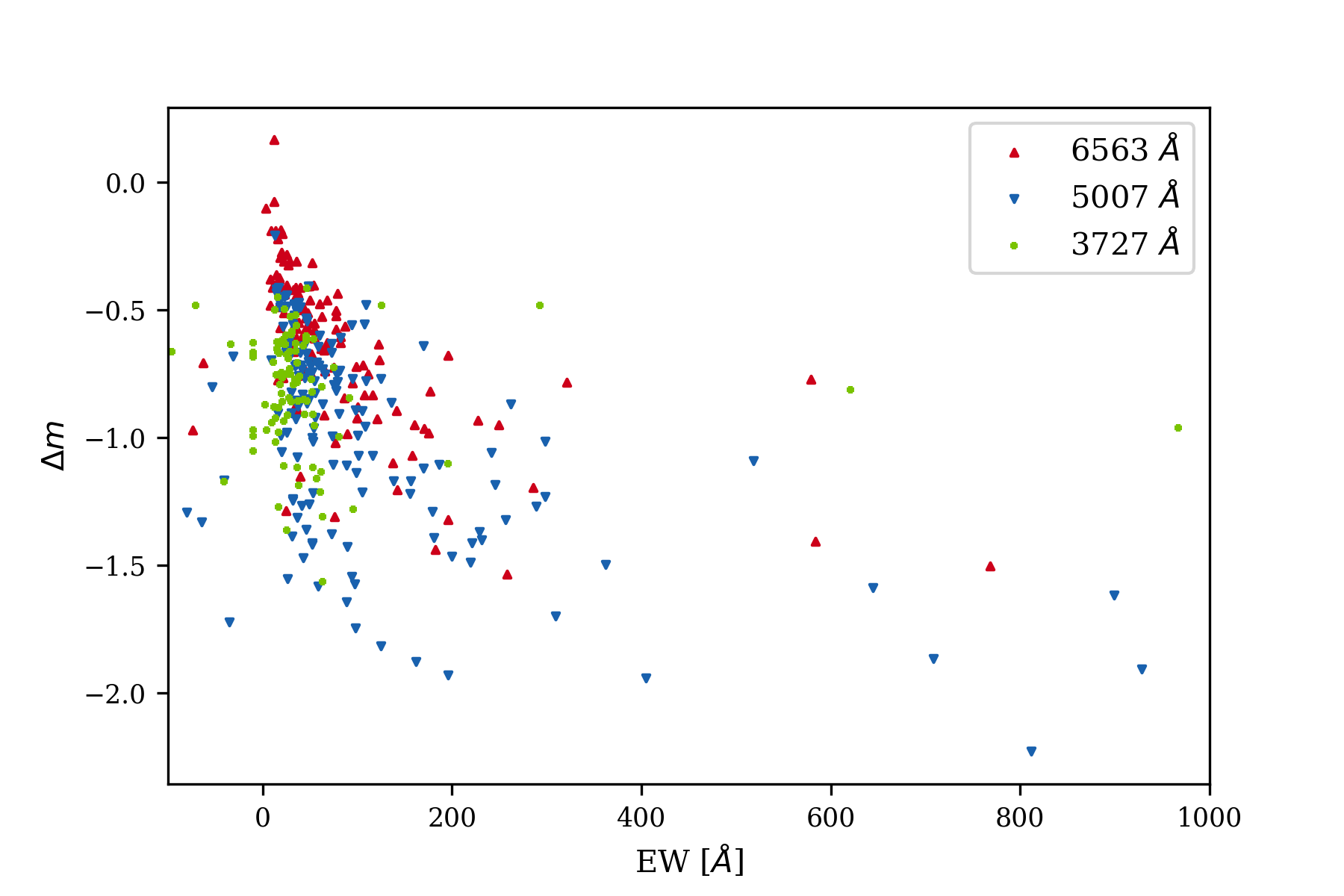}
    \caption{Shown here is the correlation between $\Delta m$ and the emission-line equivalent widths measured from our spectra (see SFACT3). Upward red triangles are objects detected via their \HA emission line, objects depicted as a blue downward triangle were detected via their \OIII emission line, and the green squares are all objects detected via their \OII emission line. The spectroscopic equivalent widths are highly uncertain, as explained in the text.   Nevertheless, the expected trend  of increasing EW with larger $-\Delta m$ is visible.}
    \label{fig:EWtrend}
\end{figure}

Despite this limitation, we can see the expected correlation between $\Delta m$ and EW in Figure~\ref{fig:EWtrend}. There is a tendency for a larger $-\Delta m$ to correlate with larger emission-line EW. This trend is true regardless of which emission line was detected in our NB filter. The figure indicates that there might be a tendency for the objects detected via $\lambda$3727 to have smaller EWs, but this could also be due to more distant and fainter objects having noisier spectra, and therefore a less well-determined continuum level. Further investigation will be conducted and addressed in future papers with a larger catalog.

\begin{figure}
    \centering
    \includegraphics[scale=1]{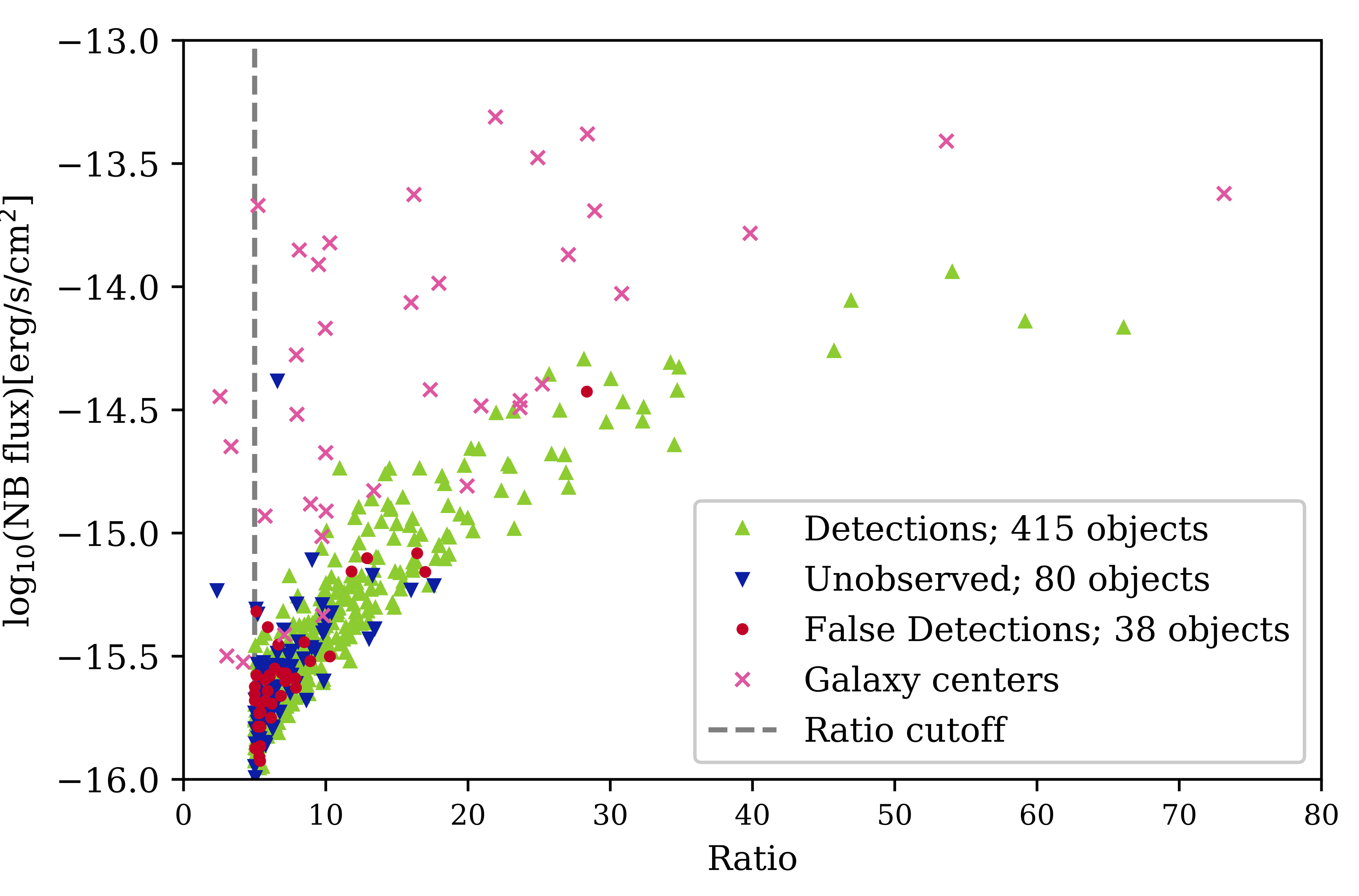}
    \includegraphics[scale=1]{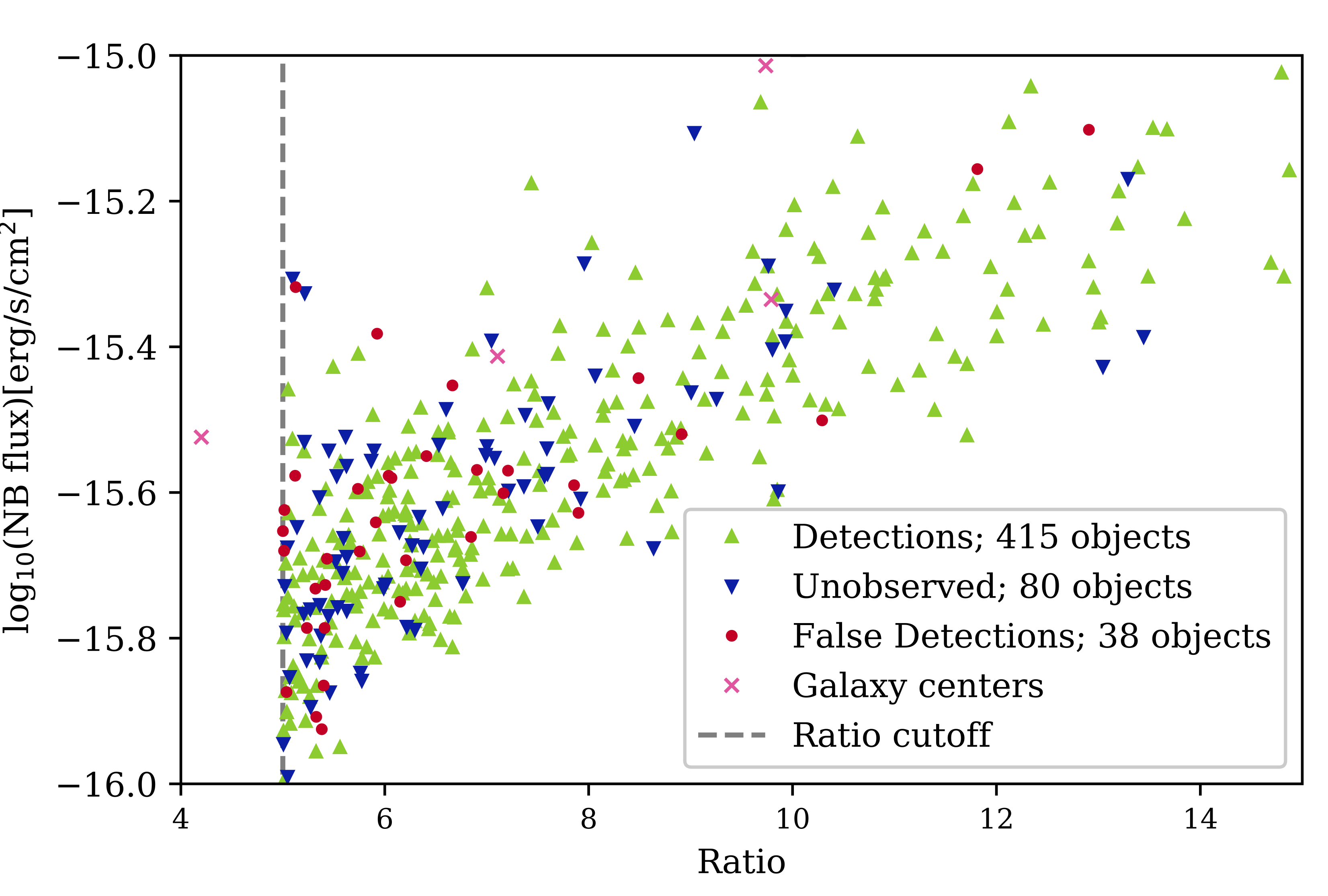}
    \caption{Shown here are the SFACT objects comparing their {\it ratio} values against their measured NB flux. The dashed vertical line shows the cutoff of objects which proceed to the next step of filtering.  Objects marked as pink crosses are galaxy centers. Blue downward triangles are candidates which do not yet have follow-up spectroscopy. Green upward triangles have been confirmed as ELGs and red circles denote objects which are confirmed to be false detections. The sample size of each is indicated in the legend. The bottom plot is a zoomed in version focusing on the location of the false detections.}
    \label{fig:RatioVnbflux}
\end{figure}

Referring back to Figure~\ref{fig:diag}, our object selection is based on {\it both} $\Delta m$ and {\it ratio}.  Hence, we next examine the relationship between the NB flux and the {\it ratio} parameter.   Since {\it ratio} is a pseudo signal-to-noise measurement, a strong signal (larger flux) should translate to a higher value of {\it ratio}.  We plot these two quantities in Figure~\ref{fig:RatioVnbflux}. The upper plot shows the full range of values for these two parameters, and reveals a strong correlation.  The only objects that deviate from the main trend are the ExtG objects, which is expected given the nature of these sources.  The lower plot is zoomed in to smaller values of {\it ratio} in order to focus on the location of the majority of our objects.  Most of the false detections are near the cutoff line, with 80\% of the false detections below {\it ratio} = 8.  Both plots show a clear correlation between {\it ratio} and the measured NB flux, as expected. 

\section{Summary and Conclusions}\label{sec:summary}

We present the initial results of the SFACT NB emission-line survey.   In the current paper we have described in detail how the imaging portion of the survey is carried out, including our observational methodology, our data processing procedures, and our object selection method.   We present our initial survey catalogs from the SFACT pilot-study fields, and present examples of detected ELGs.

By using the WIYN 3.5m telescope and ODI camera, we make good use of the wide field of view to create science fields with robust image quality across the full field of the camera. WIYN also regularly achieves sub-arcsecond seeing and has an excellent light grasp, allowing us to detect faint objects. We create a stacked master image of the three custom NB filters and the three SDSS-like BB filters. This master image gives us the depth to detect very faint objects.

We detail the procedure used to discover potential ELGs in our NB images. We search for objects using the six-filter, deep master image and then use preliminary photometry to identify those candidates which have an excess of NB flux. Our software detects candidates with significant excess flux in the NB images compared to the flux in the corresponding continuum image. These candidates are visually inspected in order to remove the image artifacts which have ELG-like signatures. Those remaining are considered SFACT candidates.

Aperture photometry is performed on all selected SFACT objects in both the BB and NB filters. SDSS stars in our images are used to calibrate the BB magnitudes and spectrophotometric stars are used to put the NB fluxes on an appropriate NB flux scale. We also demonstrate that, due to the depth of our images and the resolution of our camera, we are able to achieve reliable photometry to fairly faint magnitudes.

The 533 SFACT sources and their properties are tabulated. In these three fields, we find a surface density of 355 emission-line objects deg$^{-2}$, offering significant improvement over previous emission-line surveys that also cover modest areas of the sky (i.e., tens of square degrees). Example candidates are shown for each of the primary emission lines (H$\alpha$, [\ion{O}{3}]$\lambda$5007, and [\ion{O}{2}]$\lambda$3727) as detected in each of our NB filters. We also present one QSO at $z>1$ which was detected via its \ion{C}{3}]$\lambda$1908 line (SFF10-NB2-C21205 in Figure~\ref{find:C21205}). These example images demonstrate the wide range of objects in the SFACT catalog. Our study is dominated by faint compact objects such as SFF10-NB2-A8098 seen in Figure~\ref{Fig:OII detections}, yet SFACT also able to detect luminous QSOs. In the local universe, SFACT also detects numerous \HII regions in large extended spirals like SFF15-NB1-A2606 in Figure~\ref{Fig:HA detections}.

The photometric and NB line flux levels found for our three survey fields also demonstrate stability and good agreement. We detect objects as faint as an r-band magnitude of 25 in each of our fields and, as Figures~\ref{fig:CompositeHists} and~\ref{fig:FilterHists} demonstrate, this is achieved in all fields and in each filter. SFACT is able to detect objects with a wide range of properties all with robust photometry. 

This paper focused on the photometric results of the SFACT pilot-study fields. The corresponding spectroscopic confirmation results are discussed in greater detail in SFACT3.

We currently have an additional 35 SFACT survey fields processed, many of which already have partially-complete spectroscopic follow-up observations. These fields have the benefit of improvements to the process based on this pilot study. With thousands of additional SFACT objects in hand, future papers will begin to analyze global properties of the growing catalog and carry out the science applications planned for SFACT, as detailed in SFACT1.

\section{Acknowledgements}
The authors are honored to be permitted to conduct astronomical research on Iolkam Du’ag (Kitt Peak), a mountain with particular significance to the Tohono O’odham.

 The authors express their appreciation to the anonymous referee who made a number of insightful suggestions that improved the quality of this paper.
 
We gratefully acknowledge the long term financial support provided by the College of Arts and Sciences at Indiana University for the operation of the WIYN Observatory. Additional funds have been provided by the Department of Astronomy and the Office of the Vice Provost for Research at Indiana University to help support this project. The  SFACT  team  wishes  to  thank the entire staff of the WIYN Observatory, whose dedication and hard work have made this survey possible. In particular, we acknowledge the contributions of Daniel Harbeck, Wilson Liu, Susan Ridgeway, and Jayadev Rajagopal. We also thank Ralf Kotulla (U.\ Wisconsin) for his development and continued support of the ODI image processing software (QuickReduce), and Arvid Gopu and Michael Young (Indiana U) for their support of the ODI Pipeline, Portal \& Archive.  And we wish to thank the WIYN telescope operators without whom there would be no data. Finally, we acknowledge the contributions made at various stages of this project by students in the Department of Astronomy at Indiana University who assisted with the data processing: Bryce Cousins, Anjali Dziarski, Sean Strunk, and John Theising.

Funding for the SDSS and SDSS-II has been provided by the Alfred P.\ Sloan Foundation, the Participating Institutions, the National Science Foundation, the U.S.\ Department of Energy, the National Aeronautics and Space Administration, the Japanese Monbukagakusho, the Max Planck Society, and the Higher Education Funding Council for England. The SDSS Web Site is {\tt http://www.sdss.org/.}

The SDSS is managed by the Astrophysical Research Consortium for the Participating Institutions. The Participating Institutions are the American Museum of Natural History, Astrophysical Institute Potsdam, University of Basel, University of Cambridge, Case Western Reserve University, University of Chicago, Drexel University, Fermilab, the Institute for Advanced Study, the Japan Participation Group, Johns Hopkins University, the Joint Institute for Nuclear Astrophysics, the Kavli Institute for Particle Astrophysics and Cosmology, the Korean Scientist Group, the Chinese Academy of Sciences (LAMOST), Los Alamos National Laboratory, the Max-Planck-Institute for Astronomy (MPIA), the Max-Planck-Institute for Astrophysics (MPA), New Mexico State University, Ohio State University, University of Pittsburgh, University of Portsmouth, Princeton University, the United States Naval Observatory, and the University of Washington.

\vspace{5mm}
\facility{WIYN:3.5m}

\vspace{5mm}
\software{{\tt IRAF}}

\bibliographystyle{aasjournal}

\end{document}